\documentclass[prl,nofootinbib,amsmath,twocolumn,notitlepage,preprintnumbers]{revtex4-1}
\usepackage{amssymb,esvect,amsmath,graphicx,latexsym,amsthm,slashed,eso-pic, tabularx, bbold}
\usepackage{hyperref}
\usepackage{enumitem}
   \DeclareMathOperator{\gev}{GeV} \DeclareMathOperator{\tev}{TeV}

\newcommand{\beq}{\begin{equation}} \newcommand{\eeq}{\end{equation}}
\newcommand{\bea}{\begin{eqnarray}} \newcommand{\eea}{\end{eqnarray}}

\def\lsim{\mathrel{\raise.3ex\hbox{$<$\kern-.75em\lower1ex\hbox{$\sim$}}}}
\def\gsim{\mathrel{\raise.3ex\hbox{$>$\kern-.75em\lower1ex\hbox{$\sim$}}}}

\newcommand{\Eq}[1]{Eq.~(\ref{#1})}

\newcommand{\be}{\begin{eqnarray}}
\newcommand{\ee}{\end{eqnarray}}

\newcommand{\benum}{\begin{enumerate}}
\newcommand{\eenum}{\end{enumerate}}
\newcommand{\bi}{\begin{itemize}}
\newcommand{\ei}{\end{itemize}}

\newcommand{\me}{\slashed{E}}

\usepackage{hyperref}
\usepackage[mathlines]{lineno}

\begin{document}

\preprint{FERMILAB-PUB-20-208-AE-T
}


\title{A Guaranteed Discovery at Future Muon Colliders}

\author{Rodolfo Capdevilla$^{a,b}$}\thanks{rcapdevilla@perimeterinstitute.ca}
\author{David Curtin$^{a}$}\thanks{dcurtin@physics.utoronto.ca}
\author{Yonatan~Kahn$^{c}$}\thanks{yfkahn@illinois.edu}
\author{Gordan~Krnjaic$^d$}\thanks{krnjaicg@fnal.gov}

\affiliation{$^a$Department of Physics, University of Toronto, Canada}
\affiliation{$^b$Perimeter Institute for Theoretical Physics, Waterloo, Ontario, Canada} 
\affiliation{$^c$Illinois Center for Advanced Studies of the Universe \& Department of Physics, University of Illinois at Urbana-Champaign, Urbana, IL USA}
\affiliation{$^d$Fermi National Accelerator Laboratory, Batavia, IL USA}

\date{\today}

\begin{abstract}
The longstanding muon $g-2$ anomaly may indicate the existence of new particles that couple to muons, which could either be light~($\lsim$~GeV) and weakly coupled, or heavy~($\gg$~100 GeV) with large couplings. 
If light new states are responsible, upcoming intensity frontier experiments will discover further evidence of new physics. 
However, if heavy particles are responsible, many candidates are beyond the reach of existing colliders.
We show that, if the $(g-2)_\mu$ anomaly is confirmed and no explanation is found at low-energy experiments, a high-energy muon collider program is guaranteed to make fundamental discoveries about our universe. 
New physics scenarios that account for the anomaly can be classified as either ``Singlet" or ``Electroweak" (EW) models, involving only EW singlets or new EW-charged states respectively. 
We argue that a TeV-scale future muon collider  will discover all possible singlet model solutions to the anomaly. 
If this does not yield a discovery, the next step would be a $\mathcal{O}(10 \tev)$ muon collider. Such a machine would either discover new particles associated with high-scale EW model solutions to the anomaly, or empirically prove that nature is fine-tuned, both of which would have profound consequences for fundamental physics.

\end{abstract}

\maketitle

\section{Introduction}
The $3.7\, \sigma$ discrepancy between the Brookhaven measurement of the muon anomalous magnetic moment $a_\mu$~\cite{Bennett_2006} and the
Standard Model (SM) prediction~\cite{Aoyama:2020ynm} is among the largest and most persistent anomalies in fundamental physics.
The latest consensus~\cite{Aoyama:2012wk,Aoyama:2019ryr,Czarnecki:2002nt,Gnendiger:2013pva,Davier:2017zfy,Keshavarzi:2018mgv,Colangelo:2018mtw,Hoferichter:2019gzf,Davier:2019can,Keshavarzi:2019abf,Kurz:2014wya,Melnikov:2003xd,Masjuan:2017tvw,Colangelo:2017fiz,Hoferichter:2018kwz,Gerardin:2019vio,Bijnens:2019ghy,Colangelo:2019uex,Blum:2019ugy,Colangelo:2014qya} gives 
\be
\label{delta-amu}
\Delta a_\mu^{\rm exp} = a^{\rm exp}_\mu - a^{\rm theory}_\mu = (2.79 \pm 0.76) \times 10^{-9}~.
\ee
 If experiments at Fermilab \cite{fienberg2019status} and J-PARC \cite{Sato:2017sdn} confirm the Brookhaven result, and if precision QCD calculations do not appreciably shift the theoretical prediction, it would 
establish the first conclusive laboratory evidence of physics beyond the SM (BSM).
 
Since the new physics contribution to $a_\mu$ is fixed by coupling-to-mass ratios, the anomaly can be reconciled either with light
weakly coupled
particles \cite{Pospelov_2009}, or with heavy particles that couple appreciably to muons~\cite{Freitas:2014pua, Calibbi:2018rzv, Lindner:2016bgg, Kowalska:2017iqv, Barducci:2018esg, Kelso:2014qka, Biggio:2014ela, Queiroz:2014zfa, Biggio:2016wyy,Agrawal:2014ufa}. If the former scenario is realized in nature, multiple fixed-target experiments are
projected to discover new physics in the decade ahead \cite{Gninenko_2015,Chen_2017,Kahn_2018,kesson2018light,Berlin_2019,Tsai:2019mtm,Ballett_2019,Mohlabeng_2019,Krnjaic_2020}.
However, if these searches ultimately report null results, the only remaining possibilities involve {\it heavy} particles.


\begin{figure*}[t!]
 \vspace{0.cm}
 \hspace{-0.65cm}
 \begin{tabular}{r}
 \includegraphics[width=0.75\textwidth]{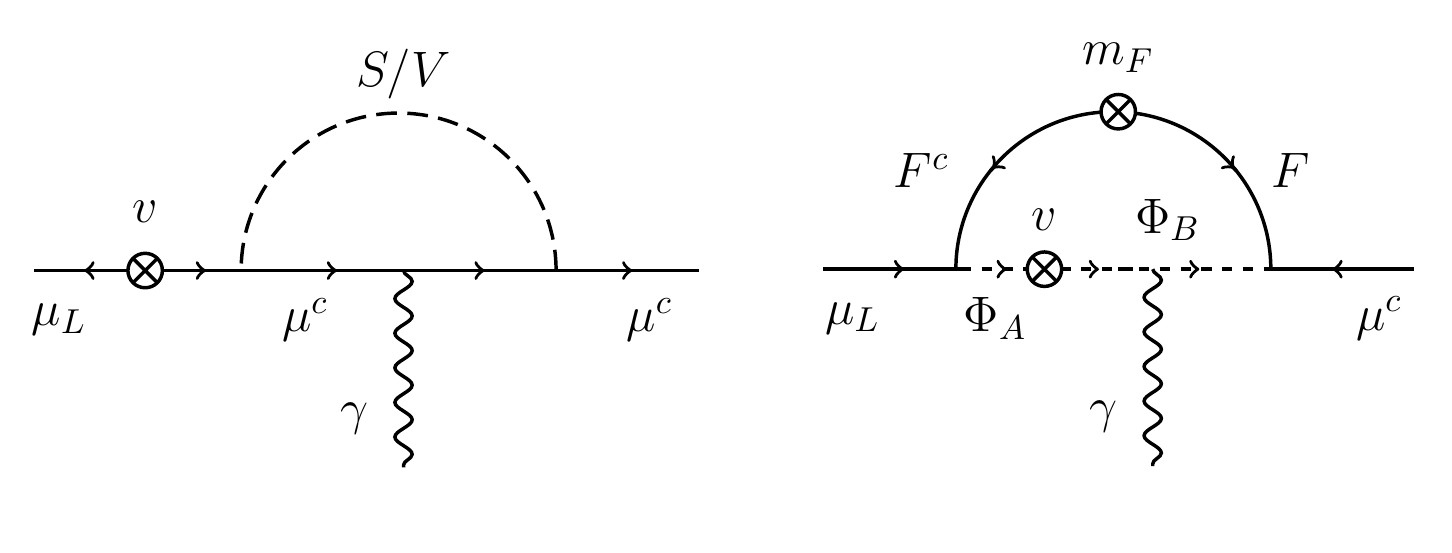}
 \vspace{-0.5cm}
\end{tabular}
\caption{Example Feynman diagrams contributing to $(g-2)_\mu$.
{\bf Left:} In models that only feature SM singlet scalars or vectors $S$ or $V$, the chirality flip and Higgs vev insertion must originate on the muon line, 
so the contribution in Eq.~(\ref{singlet-contribution}) implies ${\cal O}(1)$ couplings for singlets at the $\sim$ TeV scale. 
{\bf Right:} In scenarios that feature SM charged states, as shown for nightmare scenario, the chirality flip and EWSB Higgs coupling insertion  can be placed on internal lines, parametrically enhancing $\Delta a_\mu$ and allowing for BSM mass scales above 10 TeV. 
}
\label{feynman}
\end{figure*}


 Heavy BSM states modify $a_\mu$ through the dimension-5 operator
 \be
 \label{leff}
 {\cal L}_{\rm eff} =  C_{\rm eff}   \frac{ v}{M^2}  ({\mu}_L  \sigma^{\nu \rho}  \mu^c )  F_{\nu \rho}   + {\rm h.c.}~,
 \ee
 where $\mu_L$ and $\mu^c$ are the two-component muon fields, $v = 246$ GeV is the SM Higgs 
 vacuum expectation value (vev), 
  $C_{\rm eff}$ is a constant, and $M$ is the BSM mass scale. Note that the symmetries of the SM already impose important constraints on this operator: the chirality structure of \Eq{leff} requires a fermion mass insertion to generate $\Delta a_\mu$,
  and reconciling the different electroweak quantum numbers of $\mu_L$ and $\mu^c$ requires an insertion of $v$. 
 All BSM scenarios that generate this interaction fall into one of two categories:
 \begin{itemize}
 \item{\bf Singlet Models:} if all new particles are neutral under the SM, the Higgs coupling insertion, and hence also the chirality flip, must arise from the small muon mass $m_\mu = y_\mu v /\sqrt{2}$, so $C_{\rm eff}  \propto y_\mu$,
 where $y_\mu$ is the Higgs-muon Yukawa coupling. For the 
 maximum couplings allowed by unitarity, explaining $\Delta a_\mu$ in \Eq{delta-amu} implies $M \lsim \tev$, see Eq.~(\ref{singlet-contribution}).

 \item{\bf Electroweak (EW) Models:} if some of the new states carry $SU(2)_L \times U(1)_Y$ quantum numbers, the chirality flip and the Higgs coupling insertion in \Eq{leff} can arise from new and potentially larger masses and couplings, allowing a BSM mass scale $M \gtrsim 10 \tev$. Importantly, these interactions may yield {\it large finite} loop contributions to the Higgs mass and muon Yukawa coupling.  \end{itemize}

\noindent For both classes of models, there is a ``worst case"  scenario
in which the new particles couple preferentially to muons and are
 maximally beyond the reach of existing experiments while still generating the required $\Delta a_\mu$.   

In this {\it \small Letter} we present a ``no-lose theorem'' for a future {\it muon collider program}: 

\begin{itemize}
\item[]
If the $(g-2)_\mu$ anomaly is due to BSM physics, a combination of fixed-target experiments and a muon collider with $\sqrt{s} \gtrsim \tev$ and $\sim$10 $\mathrm{ab}^{-1}$ of luminosity will be able to discover all explanations for the anomaly involving only SM singlet fields. If no new particles are found, a higher-energy muon collider with $\sqrt{s} \sim 40 - 60 \tev$ would then be guaranteed to discover the heavy states in EW models with sizable couplings that generate $\Delta a_\mu$, or empirically prove that nature (specifically the Higgs and muon mass) is fine-tuned. If the latter is true, the BSM states generating $\Delta a_\mu$ have to have several very large couplings, and still be lighter than $\sim 100 \tev$ due to perturbative unitarity bounds. Such states would be discoverable at some future facility.
\end{itemize}
In our no-lose theorem we assume the validity of quantum field theory, so it is understood that a violation of perturbative unitarity would also be a signature of (possibly strongly-coupled) new physics with BSM states below 100 TeV.

\section{ Singlet Models}

\noindent  If the BSM states  are all EW singlets, their 
 masses do not arise from electroweak symmetry breaking (EWSB), so the chirality flip (and hence the Higgs vev insertion) in Eq.~(\ref{leff}) originates on the muon line, as shown in Fig. \ref{feynman} (left). 
  Here
  $C_{\rm eff} \sim g^2 y_\mu$, where $g$ is the singlet-muon coupling.
  These models for $g-2$ must involve at least
 one new particle coupled to  the muon,
 \be
 \label{singlet-int}
  g_S S \bar \mu \mu ~~ ,~~~
g_V V_\nu \bar \mu\gamma^\nu \mu  ~~,
  \ee
   where $S/V$ is a scalar/vector (axial or pseudoscalar couplings give the wrong sign $\Delta a_\mu$) 
and parametrically
 \be
 \label{singlet-contribution}
\Delta a_\mu \sim \frac{g^2 m_\mu^2 }{12 \pi^2 M^2}\sim  10^{-9}  g^2   \left( \frac{300 \,\rm GeV}{  M} \right)^{2},
 \ee
  where we have taken the $M \gg m_\mu$ limit \cite{Pospelov_2009,Chen_2017}.
   Thus, singlets near the weak scale must have $ \sim {\cal O}$(1) couplings to yield $\Delta a_\mu \sim 10^{-9}$ in \Eq{delta-amu}
   and the masses are bounded by $M \lesssim$ 2 TeV to satisfy unitarity bounds which require $g_{S/V} \lesssim \sqrt{4\pi}$.
   
    In what follows, we assume that the singlet $S$ or $V$ couples to the muon as in \Eq{singlet-int} 
   with sufficient strength to resolve the $\Delta a_\mu$ anomaly. We find that for all viable masses
    and decay channels, low energy experiments will test all singlet candidates below $\lesssim$ few GeV, and 
an appropriate muon collider can test the remaining heavy singlets in a model independent
fashion. 

\subsection*{Light Singlets}
Although there are many experiments designed to probe light, 
singlet particles responsible for $\Delta a_\mu$ (see \cite{Battaglieri:2017aum} for a review),
most candidates are already excluded based on how they couple
to light SM particles. Nearly all vector bosons from anomaly-free $U(1)$ SM 
gauge extensions (e.g. $B-L$) are ruled out as 
explanations for the $\Delta a_\mu$ anomaly \cite{Bauer_2018};
the only exception is 
a gauged $L_\mu - L_\tau$ gauge boson, which remains viable
for $m_V \sim 10-200$ MeV \cite{TheBABAR:2016rlg,Escudero:2019gzq}, but  will be fully tested 
with upcoming kaon decay \cite{Krnjaic:2019rsv} and muon trident searches \cite{Ballett_2019}.
Light scalars that couple preferentially to muons can still be 
 viable depending on their dominant decay modes and lifetimes~\cite{Chen_2017}.
  
 Proposed low-energy
 muon beam experiments
 can likely test all remaining $\Delta a_\mu$ candidates below the few-GeV scale 
\cite{Gninenko_2015,Chen_2017,Kahn_2018,kesson2018light,Berlin_2019,Tsai:2019mtm,Ballett_2019,Mohlabeng_2019,Krnjaic_2020, Galon:2019owl, Janish:2020knz}.
In particular, 
the proposed NA64$\mu$ \cite{Gninenko_2015,Gninenko:2019qiv} and M$^3$ \cite{Kahn_2018}  experiments are projected to cover all invisibly decaying
singlet $\Delta a_\mu$ candidates lighter than a few GeV. These  concepts can likely be modified to also test 
visibly decaying singlets produced in muon fixed-target interactions, such as a muon 
beam variation on the HPS experiment \cite{Celentano:2014wya}. Combined, these
approaches would leave no room for sub-GeV singlets that explain $\Delta a_\mu$. (Small model dependent gaps may remain for singlets that decay semi-visibly, but 
 these typically within reach of various future experiments \cite{Mohlabeng_2019}; 
 we address this possible loophole in future work~\cite{muonbible}.)

\subsection{Heavy Singlets}

Comprehensively probing all singlet candidates heavier than a few GeV that resolve the $\Delta a_\mu$ anomaly requires a muon 
collider. 
At such machines, independently of how the singlet decays, its presence  introduces
an irreducible virtual correction to muonic Bhabha scattering $\mu^+\mu^- \to \mu^+ \mu^-$. If $M \gtrsim$ 100 GeV, resolving $\Delta a_\mu$ requires 
$g \gtrsim e$, so the BSM contribution dominates and discovery is trivial. 
In the opposite regime, SM/BSM interference is the dominant signal contribution, so for $\sqrt{s} \gg M$ the BSM cross section scales as
\be
\sigma^{\rm int}_{\mu\mu \to \mu\mu} \sim \frac{g^2 \alpha }{4\pi s } \sim 2 \, {\rm fb}\,  \left( \frac{g}{10^{-2}}\right)^2  \left( \frac{\rm 100 \, \rm GeV}{\sqrt{s}}\right)^2~.
\ee
The SM Bhabha cross section scales as $ \alpha^2/s \sim 0.2 \, {\rm pb} \,({\rm 100\, GeV}/\sqrt{s})^2$, so a $5\sigma$ discovery requires a luminosity
\be
\label{lumicrit}
L \sim 10\, {\rm ab}^{-1}\, \left(  \frac{10^{-2}}{g}  \right)^4 \left(  \frac{\sqrt{s}}{ \rm 100 \, GeV}  \right)^2~,
\ee
which suffices to cover the  ``worst case" singlet scenario for $M \sim$ 5 GeV,
just beyond the kinematic reach of muon beam fixed target experiments \cite{Gninenko_2015,Gninenko:2019qiv,Kahn_2018}. Note that the 
$g^{-4}$ scaling in \Eq{lumicrit} guarantees that all heavier $\Delta a_\mu$ candidate singlets can be discovered with even less luminosity; from \Eq{singlet-contribution}, $g$ must be even larger to resolve the anomaly at higher mass. Note that our analysis here conservatively 
relies solely on singlet exchange in muonic Bhabha scattering, which makes no assumptions about how the singlet decays. 

Thus, a combination of fixed target and muon collider searches at various $\sqrt{s}$ can cover all remaining singlet models for $\Delta a_\mu$ up to the unitarity limit at $M \sim$ TeV. Note that if there are $N_{\rm BSM} >1$ singlets coupled to the muon, then the highest  mass scale  compatible with unitarity  may increase with $\sqrt{N_{\rm BSM}}$, which would not significantly change our conclusions for $N_{\rm BSM} \lesssim 10$.

 \section{Electroweak Models} 
 
\noindent Much higher BSM mass scales are possible if the new states carry EW quantum numbers, which allows the chirality flip and/or $v$ insertion in Eq.~(\ref{leff}) to arise from heavy BSM states, yielding
\be
 \label{doublet-contribution}
\Delta a_\mu \sim  \frac{ y^3 m_\mu v }{8 \pi^2 M^2} \sim 10^{-9} C_{\rm eff}  \left( \frac{20 \,\rm TeV}{  M} \right)^{2}
 \ee
with $C_\mathrm{eff} \sim 1$. Here $y$ is a generic trilinear coupling, see Fig. \ref{feynman} (right). 
Although there are many possibilities for such BSM models, 
we are interested in the minimum energy and luminosity a muon collider must have for a \emph{ guaranteed discovery of the highest-scale BSM models generating $\Delta a_\mu$}.  Thus, we need only study the minimal simplified models that generate the \emph{largest possible} 1-loop  $\Delta a_\mu$ contributions for a given BSM mass scale $M_{\rm BSM}$. 
Note that we restrict to 1-loop contributions because 2- and higher-loop contributions will give a lower BSM mass scale.
This leads us to consider some of the models previously studied in Refs.~\cite{Calibbi:2018rzv, Agrawal:2014ufa, Kowalska:2017iqv, Barducci:2018esg, Calibbi:2019bay,Calibbi:2020emz,Crivellin:2018qmi}, though in a very different context. Our approach is also different from previous attempts to define simplified model dictionaries for generating $\Delta a_\mu$~\cite{Freitas:2014pua, Calibbi:2018rzv, Lindner:2016bgg, Kowalska:2017iqv, Barducci:2018esg, Kelso:2014qka, Biggio:2014ela, Queiroz:2014zfa, Biggio:2016wyy},
since we identify a single ``nightmare scenario'' that acts as a synecdoche for all possible perturbative models, in order to determine the highest possible BSM mass scale.

\subsection{Simplified Nightmare Scenario for $\Delta a_\mu$}

The necessary ingredients are: (1)~at least 3 BSM fields, including at least one boson and one fermion; (2)~a pair of those fields mixes via a Higgs coupling after EWSB; (3)~all new fermions are vector-like to maximize allowed masses; (4)~all new scalars with EW charges do not acquire vacuum expectation values, since for~$>$~TeV scales, any EWSB vev exceeds the measured $v \approx 246 \gev$ for perturbative scalar self-couplings. 
We also focus on the most experimentally pessimistic case in which BSM states only couple to the SM 
through their muonic interactions. 

Thus, we define a model with a single vector-like fermion pair $F/F^c$ in $SU(2)_L$ representation $R^F$ with hypercharge $Y^F$, and two complex scalars $\Phi_A, \Phi_B$ in $SU(2)_L$ representations $R^A, R^B$ with hypercharges $Y^A, Y^B$:
\be
\label{SLR}
\mathcal{L} &\supset& -y_1 F^c L_{(\mu)}  \Phi_A^*  - y_2 F \mu^c  \Phi_B 
- \kappa H \Phi_A^* \Phi_B 
\nonumber \\  &&
- m^2_A |\Phi_A|^2 - m^2_B |\Phi_B|^2 - m_F F F^c
 + h.c. \ .
\ee
Here $y_1, y_2$ are new Yukawa couplings and $\kappa$ is a trilinear coupling with dimensions of mass.
$L_{(\mu)}$ and $\mu^c$ are the left- and right-handed second-generation SM lepton fields, and $H$ is the Higgs doublet. 
The choices of representations must satisfy $R^A \otimes R^F \otimes \mathbf{2} \supset \mathbf{1}$, $R^B = R^F$ and $Y^A = -\frac{1}{2} - Y^F$, $Y^B = -1 - Y^F$.
We also consider the trivial generalization where there are $N_{\rm BSM}$ copies of the above fields (i.e.\ BSM flavors) contributing to $\Delta a_\mu$. 

There are other representative model classes that satisfy the above requirements for ``nightmare scenarios," including  (1) scenarios with two fermions and a scalar (instead of two scalars and a fermion as above); (2) models with vectors instead of scalars;  or  (3) cases involving Majorana fermions or real scalars. We analyzed these cases and checked that all such variations yield smaller $\Delta a_\mu$ contributions, and hence require lower BSM mass scales to explain the anomaly. It is also possible for additional couplings to be present in \Eq{SLR}, but they do not contribute to $\Delta a_\mu$ and would not significantly change our analysis, except by possibly introducing additional collider signatures; we conservatively ignore these other possibilities. 

We calculate the $\Delta a_\mu$ contribution of our nightmare scenario, shown in Fig.~\ref{feynman} (right), and reproduce results in the literature~\cite{Freitas:2014pua, Calibbi:2018rzv}. The chirality flip comes from the vector-like fermion mass, while the $v$ insertion is due to $\Phi_A$-$\Phi_B$ mixing, giving $\Delta a_\mu \sim y_{1} y_{2}  \kappa v  m_\mu /M_{\rm BSM}^3$. 
Note that $\Delta a_\mu$ is now controlled by three BSM couplings $y_1, y_2, \kappa$ that can be relatively large.

\subsection{Highest possible BSM mass for $\Delta a_\mu$}

For our nightmare scenario, we study all possible representations in which every new particle 
is a  singlet, doublet or triplet under $SU(2)_L$ (similar to~\cite{Lindner:2016bgg, Kowalska:2017iqv})
and has electric charge $|Q| \leq 2$. 
Generalizing our results is straightforward, though we do not expect higher 
representations to change our conclusions, since we find that the largest BSM masses are allowed for the ``smallest'' models with two singlets and one doublet.

In each of these cases, for a given choice of mass parameters $m_A, m_B$, we calculate $\Delta a_\mu$ at 1-loop and find the highest possible mass $m_F$ that yields $\Delta a_\mu = 2.8 \times 10^{-9}$ for allowed values of  $y_1, y_2, \kappa$. 
 The couplings are restricted to satisfy perturbative unitarity constraints (we derive them in detail in \cite{muonbible} similar to~\cite{Lee:1977yc,Haber:1994pe,Castillo:2013uda,Queiroz:2014zfa,Goodsell:2018tti,Biondini:2019tcc}),
 which are  $|y_{1,2}| \leq \sqrt{16 \pi}$, $|\kappa| < \kappa_{\rm max}$, where $\kappa_{\rm max} \sim d m_A m_B/v$ is a function of BSM mass parameters and $d \sim \mathcal{O}(0.1 - 1)$
if there is large hierarchy between $m_A$ and $m_B$, asymptoting to $d \ll 1$ as $m_A \to m_B$.

While unitarity dictates the only physics upper bound on BSM masses, it is important to consider the naturalness of the model as well, since generating the required $\Delta a_\mu$ with maximally heavy BSM masses introduces \emph{two hierarchy problems}, due to large finite loop corrections to the Higgs mass and the muon Yukawa:
\beq
\Delta m_H^2  \sim  - \frac{\kappa^2}{16 \pi^2}
\ \ \  , \ \   
\Delta y_\mu   \sim   \frac{y_1 y_2 }{16 \pi^2} \  \frac{\kappa\, m_F}{M^2} \ ,
\eeq
where $M^2$ is a combination of $m_A^2, m_B^2, m_F^2$. 
This is not surprising for the Higgs, which has a well-known quadratic sensitivity to new physics. It is interesting, however, that the muon mass becomes technically un-natural in this nightmare scenario (and its variations mentioned above), since 
the 1-loop correction to $y_\mu$ is no longer proportional to $y_\mu$ itself, due to the shared chiral symmetry between muons and the new heavy fermions in the limit where \emph{both} are massless.

These (two) hierarchy problems are not like the (single) hierarchy problem of the SM, which arises due to well-motivated but still hypothetical contributions in the far UV. Rather, they concretely arise from heavy BSM particles solving the $(g-2)_\mu$ anomaly, and represent an explicit tuning of Lagrangian parameters due to finite, calculable contributions within the theory. 
Experimental determination of their high mass scale would therefore constitute \emph{empirical proof that nature is tuned!}
(A similar observation was recently made in connection with electron EDM measurements~\cite{Cesarotti:2018huy} and also in~\cite{Calibbi:2020emz}.)
This is analogous to discovering e.g. split supersymmetry~\cite{Giudice:2004tc,ArkaniHamed:2004yi}, where the lightest new physics states are heavy and couple to the Higgs, except that to solve the $(g-2)_\mu$ anomaly with such heavy states, the muon mass must be tuned and technically un-natural as well.

\begin{figure*}[t!]
 \vspace{0.cm}
 \hspace{-0.65cm}
 \begin{tabular}{ccc}
 \includegraphics[width=0.35\textwidth]{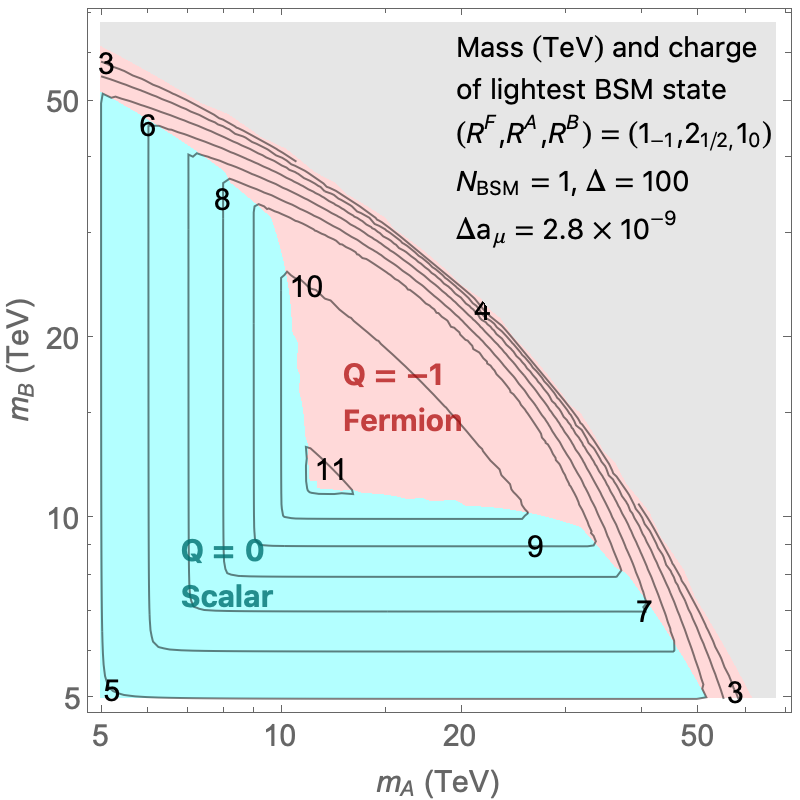}
 ~~~~
 \includegraphics[width=0.35\textwidth]{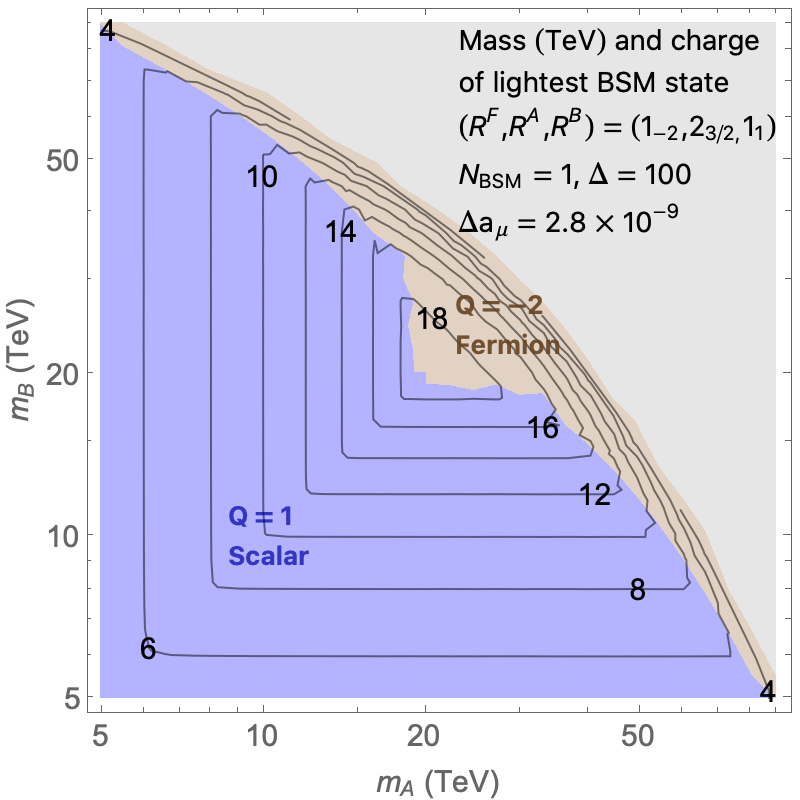}
\end{tabular}
\caption{
Contours show the mass (TeV) of the lightest BSM state in two EW nightmare scenarios with a single BSM flavor
$(R^F_{\ Y^F}, R^A_{\ Y^A}, R^B_{\ Y^B}) =  (\mathbf{1}_{-1}, \mathbf{2}_{\frac{1}{2}}\mathbf{1}_{0})$ (left) or $(\mathbf{1}_{-2}, \mathbf{2}_{\frac{3}{2}}, \mathbf{1}_{1})$ (right), 
as a function of the scalar mass parameters $(m_A, m_B)$. %
The colored regions are labeled with their lightest BSM state.
For each $(m_A, m_B)$, we found the largest possible fermion mass $m_F$ and couplings  $y_1, y_2, \kappa$ 
that generate $\Delta a_\mu = 2.8 \times 10^{-9}$ without exceeding Higgs and muon mass tuning $\Delta = 100$. 
In the gray regions, the required $\Delta a_\mu$ cannot be generated. 
}
\label{results}
\end{figure*}

We therefore also ask how heavy the BSM states can be if we impose a conservative naturalness constraint on the new couplings by requiring 
\beq
\Delta \equiv \mathrm{max}\left(\frac{\Delta m_H^2}{m_H^2}, \frac{\Delta y_\mu}{y_\mu}\right) < 100 \ .
\eeq
This  corresponds to percent-level tuning in \emph{both} Higgs and muon masses, since maximizing all couplings relevant for $\Delta  a_\mu$ saturates both tuning bounds.
Fig.~\ref{results} shows our results for the two scenarios with the highest mass of the lightest BSM state, in the case where that state can be neutral (left) or always has to be charged (right).
For $N_{\rm BSM} > 1$, one expects higher BSM masses to be possible since there are more contributions to $\Delta a_\mu$, but each flavor also contributes to the  $\Delta m_{H}^2, \Delta  y_\mu$ corrections. 
This lowers the maximum allowed size of the couplings for a given tuning, resulting in the largest possible BSM masses depending only very weakly on the number of flavors. Empirically, we find that the mass upper bounds scale as $\sim N_{\rm BSM}^{1/6}$.\footnote{For some choices of BSM gauge representations, the $N_\mathrm{BSM}$ dependence of the heaviest BSM mass cancels completely, but this does not affect the overall upper bound on the mass of charged particles.}

\emph{Therefore, given the different possible charge assignments in our nightmare scenarios, and allowing for up to percent-level tuning of both $m^2_H$ and $m_\mu$, 
we find that
$\Delta a_\mu = 2.8 \times 10^{-9}$ with one (10) BSM flavor(s) predicts 
at least one new charged particle lighter than 20 (30) TeV. 
In some cases, the lightest BSM state is a SM-singlet, in which case its mass must be below 12 TeV.}

The maximum allowed BSM masses are somewhat higher if only unitarity constraints are imposed, allowing all BSM couplings to be maximally large. Neutral or charged particles must then still be lighter  than 50/100 TeV for $N_{\rm BSM} = 1$, with maximum masses scaling as $\sqrt{N_{\rm BSM}}$.

\subsection{Muon Collider Signatures}

Based on our findings for the EW nightmare scenarios, we now argue that a $\sqrt{s} \sim 40-60$~TeV muon collider with sufficient luminosity to study benchmark SM processes at this energy will be able to  discover \emph{either} heavy BSM particles generating $\Delta a_\mu$, \emph{or} prove that nature is fine-tuned by a process of elimination (assuming lower-scale singlet model solutions to the anomaly have been excluded by that point).

High-scale EW model solutions to the $(g-2)_\mu$ anomaly require charged states with masses below $\sim 20-30$ TeV, for $N_{\rm BSM} = 1 - 10$. This sets the minimum energy of the muon collider required to guarantee the pair-production of BSM states at $\sqrt{s} \sim 40 - 60 \tev$ (unless one allows for extremely high numbers of BSM flavors).
A heavy charged state $X$ can be pair-produced in Drell-Yan processes independent of its direct couplings to muons, with a pair production cross section similar to SM EW $2\to 2$ processes above threshold, $\sigma_{XX} \sim \mathrm{fb} \, (10 \tev/\sqrt{s})^2$~\cite{Delahaye:2019omf}. 
Since such states are visible if they are detector-stable, or have to decay to visible SM final states, conclusive discovery of such heavy new states should be possible regardless of their detailed phenomenology.
Of course, a high-energy electron or much higher-energy proton collider could also produce these states. This would likely be even more technically daunting than a muon collider at this energy (see also~\cite{Costantini:2020stv}).

While at least one of the charged states can always be produced in EW processes, in some cases the lightest state $X_0$ is a SM singlet. 
This singlet has sizable direct couplings  $y_{1,2}$ to muons, meaning only a muon collider is guaranteed to produce $X_0$ via $t$-channel exchange of a heavy charged state with cross section $\sigma \sim y_{1,2}^2/(4 \pi s) \sim \mathcal{O}(10 \ \mathrm{fb})$ for $y_{1,2} \sim \mathcal{O}(1)$. 
(Smaller couplings can generate $\Delta a_\mu$ only if all the BSM masses are significantly below our upper bounds, in which case many new states become discoverable.)  
We perform basic signal and background estimates using MadGraph5~\cite{Alwall:2014hca}, leaving a more detailed collider study for future work \cite{muonbible}.

If $X_0$ decays visibly inside the detector, it can be discovered in similar searches as for the charged states.
If it is detector-stable, discovery has to rely on $X_0$ pair production in association with a photon, leading to a $\gamma + \me$ signal with cross sections in the $1 - 10$ fb range for $p_\gamma \gtrsim \tev$. 
The main SM backgrounds are $\mu^+ \mu^- \to \gamma(Z \to \bar \nu \nu)$, which is easily vetoed since it is dominated by on-shell $Z$-production with $p_\gamma \approx \sqrt{s}/2$, and vector boson fusion (VBF) processes like $\mu^+ \mu^- \to \bar \nu \nu \gamma$ with $t$-channel $W$-exchange. 
The latter has a large cross section $\sim \mathrm{pb}$, since
SM EW production processes via VBF are greatly enhanced at high-energy muon colliders~\cite{Costantini:2020stv}, but the associated photon is relatively soft, so imposing the $p_\gamma \gtrsim \tev$ cut reduces the total background rate to $\sim \mathcal{O}(10)$~fb, meaning discovery is possible with 10 - 100 $\mathrm{ab}^{-1}$ of luminosity depending on the BSM couplings.

\section{Conclusion}
 
  The search for physics beyond the SM is one of the key pursuits of high-energy physics. Unlike other possible sources of BSM physics, such as dark matter or a solution to the strong-CP problem, a BSM explanation for the muon anomalous magnetic moment \emph{requires} new states with couplings to SM particles and a mass scale bounded from above. In this \emph{Letter}, we have outlined a model-independent search strategy which, assuming the $a_\mu$ anomaly is genuine, is \emph{guaranteed} to discover new physics in the same way that the LHC was guaranteed to discover new physics related to electroweak symmetry breaking.

 It has recently been argued that a muon collider with $\sqrt{s} \sim 10$ TeV and  integrated luminosity of $\sim$ 10 ab$^{-1}$  in $\sim 3$ years of running is a potentially achievable  \cite{Delahaye:2019omf,Strategy:2019vxc}. 
 A facility ramping up to this energy would be able to discover all singlet model explanations for the $(g-2)_\mu$ anomaly, and probe the most theoretically motivated (i.e. natural) parameter space of all the EW models as well. 
 Furthermore, because muon colliders utilize the full CM energy towards producing on-shell states, $\sqrt{s} = 10$ TeV
corresponds to an equivalent hadron collider CM energy  of $\approx 220$ TeV \cite{Costantini:2020stv}, thereby
enabling a broad exploration of SM and BSM physics at unprecedented energy scales, including various 
electroweak dark matter scenarios \cite{DiLuzio:2018jwd,Costantini:2020stv}, and precise measurement of the Higgs quartic coupling \cite{chiesa2020measuring}.

In future work~\cite{muonbible}, we will elaborate on the phenomenology of the various possible Electroweak models, the precise unitarity bounds for each coupling from each scattering channel, and conduct detailed studies of the dominant signatures for each model as a function of muon collider energy and luminosity that expand on our initial estimates. We will also discuss how departures from the starkly minimal assumptions we have made would \emph{lower} the BSM mass scale due to e.g. flavor constraints on related couplings~\cite{Lindner:2016bgg, Batell:2017kty}.
 The goal of the present \emph{Letter} and this longer follow-up compendium will be to set the stage for a muon collider that would pay concrete physics dividends in the coming decades.

\bigskip
\begin{acknowledgments}  
\textbf{Acknowledgments.} We are grateful to Asimina Arvanitaki, Doug Berry, Zackaria Chacko, Matteo Cremonesi,  Aida El-Khadra,  Allison Hall,  Dan Hooper, Bo Jayatilaka, Jessie Shelton, Daniel Stolarski, Raman Sundrum and Nhan Tran for helpful conversations. 
All the authors would like to express their deep appreciation for the wanton travel and social interaction of the pre-pandemic times, which allowed the idea for this work to be conceived.
The work of DC and RC was supported in part by a Discovery Grant from the Natural Sciences and Engineering Research Council of Canada, and by the Canada Research Chair program.
The work of RC was supported in part by the Perimeter Institute for Theoretical Physics (PI). Research at PI is supported in part by the Government of Canada through the Department of Innovation, Science and Economic Development Canada and by the Province of Ontario through the Ministry of Colleges and Universities.
The work of GK is supported by the Fermi Research Alliance, LLC under Contract No. DE-AC02-07CH11359 with the U.S. Department of Energy, Office of Science, Office of High Energy Physics. 
\end{acknowledgments}

\bibliography{NoLose}

\begin{thebibliography}{72}%
\makeatletter
\providecommand \@ifxundefined [1]{%
 \@ifx{#1\undefined}
}%
\providecommand \@ifnum [1]{%
 \ifnum #1\expandafter \@firstoftwo
 \else \expandafter \@secondoftwo
 \fi
}%
\providecommand \@ifx [1]{%
 \ifx #1\expandafter \@firstoftwo
 \else \expandafter \@secondoftwo
 \fi
}%
\providecommand \natexlab [1]{#1}%
\providecommand \enquote  [1]{``#1''}%
\providecommand \bibnamefont  [1]{#1}%
\providecommand \bibfnamefont [1]{#1}%
\providecommand \citenamefont [1]{#1}%
\providecommand \href@noop [0]{\@secondoftwo}%
\providecommand \href [0]{\begingroup \@sanitize@url \@href}%
\providecommand \@href[1]{\@@startlink{#1}\@@href}%
\providecommand \@@href[1]{\endgroup#1\@@endlink}%
\providecommand \@sanitize@url [0]{\catcode `\\12\catcode `\$12\catcode
  `\&12\catcode `\#12\catcode `\^12\catcode `\_12\catcode `\%12\relax}%
\providecommand \@@startlink[1]{}%
\providecommand \@@endlink[0]{}%
\providecommand \url  [0]{\begingroup\@sanitize@url \@url }%
\providecommand \@url [1]{\endgroup\@href {#1}{\urlprefix }}%
\providecommand \urlprefix  [0]{URL }%
\providecommand \Eprint [0]{\href }%
\providecommand \doibase [0]{http://dx.doi.org/}%
\providecommand \selectlanguage [0]{\@gobble}%
\providecommand \bibinfo  [0]{\@secondoftwo}%
\providecommand \bibfield  [0]{\@secondoftwo}%
\providecommand \translation [1]{[#1]}%
\providecommand \BibitemOpen [0]{}%
\providecommand \bibitemStop [0]{}%
\providecommand \bibitemNoStop [0]{.\EOS\space}%
\providecommand \EOS [0]{\spacefactor3000\relax}%
\providecommand \BibitemShut  [1]{\csname bibitem#1\endcsname}%
\let\auto@bib@innerbib\@empty
\bibitem [{\citenamefont {Bennett}\ \emph {et~al.}(2006)\citenamefont
  {Bennett}, \citenamefont {Bousquet}, \citenamefont {Brown}, \citenamefont
  {Bunce}, \citenamefont {Carey}, \citenamefont {Cushman}, \citenamefont
  {Danby}, \citenamefont {Debevec}, \citenamefont {Deile}, \citenamefont
  {Deng},\ and\ \citenamefont {et~al.}}]{Bennett_2006}%
  \BibitemOpen
  \bibfield  {author} {\bibinfo {author} {\bibfnamefont {G.~W.}\ \bibnamefont
  {Bennett}}, \bibinfo {author} {\bibfnamefont {B.}~\bibnamefont {Bousquet}},
  \bibinfo {author} {\bibfnamefont {H.~N.}\ \bibnamefont {Brown}}, \bibinfo
  {author} {\bibfnamefont {G.}~\bibnamefont {Bunce}}, \bibinfo {author}
  {\bibfnamefont {R.~M.}\ \bibnamefont {Carey}}, \bibinfo {author}
  {\bibfnamefont {P.}~\bibnamefont {Cushman}}, \bibinfo {author} {\bibfnamefont
  {G.~T.}\ \bibnamefont {Danby}}, \bibinfo {author} {\bibfnamefont {P.~T.}\
  \bibnamefont {Debevec}}, \bibinfo {author} {\bibfnamefont {M.}~\bibnamefont
  {Deile}}, \bibinfo {author} {\bibfnamefont {H.}~\bibnamefont {Deng}}, \ and\
  \bibinfo {author} {\bibnamefont {et~al.}},\ }\href {\doibase
  10.1103/physrevd.73.072003} {\bibfield  {journal} {\bibinfo  {journal}
  {Physical Review D}\ }\textbf {\bibinfo {volume} {73}} (\bibinfo {year}
  {2006}),\ 10.1103/physrevd.73.072003}\BibitemShut {NoStop}%
\bibitem [{\citenamefont {Aoyama}\ \emph {et~al.}(2020)\citenamefont {Aoyama}
  \emph {et~al.}}]{Aoyama:2020ynm}%
  \BibitemOpen
  \bibfield  {author} {\bibinfo {author} {\bibfnamefont {T.}~\bibnamefont
  {Aoyama}} \emph {et~al.},\ }\href@noop {} {\  (\bibinfo {year} {2020})},\
  \Eprint {http://arxiv.org/abs/2006.04822} {arXiv:2006.04822 [hep-ph]}
  \BibitemShut {NoStop}%
\bibitem [{\citenamefont {Aoyama}\ \emph {et~al.}(2012)\citenamefont {Aoyama},
  \citenamefont {Hayakawa}, \citenamefont {Kinoshita},\ and\ \citenamefont
  {Nio}}]{Aoyama:2012wk}%
  \BibitemOpen
  \bibfield  {author} {\bibinfo {author} {\bibfnamefont {T.}~\bibnamefont
  {Aoyama}}, \bibinfo {author} {\bibfnamefont {M.}~\bibnamefont {Hayakawa}},
  \bibinfo {author} {\bibfnamefont {T.}~\bibnamefont {Kinoshita}}, \ and\
  \bibinfo {author} {\bibfnamefont {M.}~\bibnamefont {Nio}},\ }\href {\doibase
  10.1103/PhysRevLett.109.111808} {\bibfield  {journal} {\bibinfo  {journal}
  {Phys. Rev. Lett.}\ }\textbf {\bibinfo {volume} {109}},\ \bibinfo {pages}
  {111808} (\bibinfo {year} {2012})},\ \Eprint {http://arxiv.org/abs/1205.5370}
  {arXiv:1205.5370 [hep-ph]} \BibitemShut {NoStop}%
\bibitem [{\citenamefont {Aoyama}\ \emph {et~al.}(2019)\citenamefont {Aoyama},
  \citenamefont {Kinoshita},\ and\ \citenamefont {Nio}}]{Aoyama:2019ryr}%
  \BibitemOpen
  \bibfield  {author} {\bibinfo {author} {\bibfnamefont {T.}~\bibnamefont
  {Aoyama}}, \bibinfo {author} {\bibfnamefont {T.}~\bibnamefont {Kinoshita}}, \
  and\ \bibinfo {author} {\bibfnamefont {M.}~\bibnamefont {Nio}},\ }\href
  {\doibase 10.3390/atoms7010028} {\bibfield  {journal} {\bibinfo  {journal}
  {Atoms}\ }\textbf {\bibinfo {volume} {7}},\ \bibinfo {pages} {28} (\bibinfo
  {year} {2019})}\BibitemShut {NoStop}%
\bibitem [{\citenamefont {Czarnecki}\ \emph {et~al.}(2003)\citenamefont
  {Czarnecki}, \citenamefont {Marciano},\ and\ \citenamefont
  {Vainshtein}}]{Czarnecki:2002nt}%
  \BibitemOpen
  \bibfield  {author} {\bibinfo {author} {\bibfnamefont {A.}~\bibnamefont
  {Czarnecki}}, \bibinfo {author} {\bibfnamefont {W.~J.}\ \bibnamefont
  {Marciano}}, \ and\ \bibinfo {author} {\bibfnamefont {A.}~\bibnamefont
  {Vainshtein}},\ }\href {\doibase 10.1103/PhysRevD.67.073006} {\bibfield
  {journal} {\bibinfo  {journal} {Phys. Rev.}\ }\textbf {\bibinfo {volume}
  {D67}},\ \bibinfo {pages} {073006} (\bibinfo {year} {2003})},\ \bibinfo
  {note} {[Erratum: Phys. Rev. {\bf D73}, 119901 (2006)]},\ \Eprint
  {http://arxiv.org/abs/hep-ph/0212229} {arXiv:hep-ph/0212229 [hep-ph]}
  \BibitemShut {NoStop}%
\bibitem [{\citenamefont {Gnendiger}\ \emph {et~al.}(2013)\citenamefont
  {Gnendiger}, \citenamefont {St{\"o}ckinger},\ and\ \citenamefont
  {St{\"o}ckinger-Kim}}]{Gnendiger:2013pva}%
  \BibitemOpen
  \bibfield  {author} {\bibinfo {author} {\bibfnamefont {C.}~\bibnamefont
  {Gnendiger}}, \bibinfo {author} {\bibfnamefont {D.}~\bibnamefont
  {St{\"o}ckinger}}, \ and\ \bibinfo {author} {\bibfnamefont {H.}~\bibnamefont
  {St{\"o}ckinger-Kim}},\ }\href {\doibase 10.1103/PhysRevD.88.053005}
  {\bibfield  {journal} {\bibinfo  {journal} {Phys. Rev.}\ }\textbf {\bibinfo
  {volume} {D88}},\ \bibinfo {pages} {053005} (\bibinfo {year} {2013})},\
  \Eprint {http://arxiv.org/abs/1306.5546} {arXiv:1306.5546 [hep-ph]}
  \BibitemShut {NoStop}%
\bibitem [{\citenamefont {Davier}\ \emph {et~al.}(2017)\citenamefont {Davier},
  \citenamefont {Hoecker}, \citenamefont {Malaescu},\ and\ \citenamefont
  {Zhang}}]{Davier:2017zfy}%
  \BibitemOpen
  \bibfield  {author} {\bibinfo {author} {\bibfnamefont {M.}~\bibnamefont
  {Davier}}, \bibinfo {author} {\bibfnamefont {A.}~\bibnamefont {Hoecker}},
  \bibinfo {author} {\bibfnamefont {B.}~\bibnamefont {Malaescu}}, \ and\
  \bibinfo {author} {\bibfnamefont {Z.}~\bibnamefont {Zhang}},\ }\href
  {\doibase 10.1140/epjc/s10052-017-5161-6} {\bibfield  {journal} {\bibinfo
  {journal} {Eur. Phys. J.}\ }\textbf {\bibinfo {volume} {C77}},\ \bibinfo
  {pages} {827} (\bibinfo {year} {2017})},\ \Eprint
  {http://arxiv.org/abs/1706.09436} {arXiv:1706.09436 [hep-ph]} \BibitemShut
  {NoStop}%
\bibitem [{\citenamefont {Keshavarzi}\ \emph {et~al.}(2018)\citenamefont
  {Keshavarzi}, \citenamefont {Nomura},\ and\ \citenamefont
  {Teubner}}]{Keshavarzi:2018mgv}%
  \BibitemOpen
  \bibfield  {author} {\bibinfo {author} {\bibfnamefont {A.}~\bibnamefont
  {Keshavarzi}}, \bibinfo {author} {\bibfnamefont {D.}~\bibnamefont {Nomura}},
  \ and\ \bibinfo {author} {\bibfnamefont {T.}~\bibnamefont {Teubner}},\ }\href
  {\doibase 10.1103/PhysRevD.97.114025} {\bibfield  {journal} {\bibinfo
  {journal} {Phys. Rev.}\ }\textbf {\bibinfo {volume} {D97}},\ \bibinfo {pages}
  {114025} (\bibinfo {year} {2018})},\ \Eprint
  {http://arxiv.org/abs/1802.02995} {arXiv:1802.02995 [hep-ph]} \BibitemShut
  {NoStop}%
\bibitem [{\citenamefont {Colangelo}\ \emph {et~al.}(2019)\citenamefont
  {Colangelo}, \citenamefont {Hoferichter},\ and\ \citenamefont
  {Stoffer}}]{Colangelo:2018mtw}%
  \BibitemOpen
  \bibfield  {author} {\bibinfo {author} {\bibfnamefont {G.}~\bibnamefont
  {Colangelo}}, \bibinfo {author} {\bibfnamefont {M.}~\bibnamefont
  {Hoferichter}}, \ and\ \bibinfo {author} {\bibfnamefont {P.}~\bibnamefont
  {Stoffer}},\ }\href {\doibase 10.1007/JHEP02(2019)006} {\bibfield  {journal}
  {\bibinfo  {journal} {JHEP}\ }\textbf {\bibinfo {volume} {02}},\ \bibinfo
  {pages} {006} (\bibinfo {year} {2019})},\ \Eprint
  {http://arxiv.org/abs/1810.00007} {arXiv:1810.00007 [hep-ph]} \BibitemShut
  {NoStop}%
\bibitem [{\citenamefont {Hoferichter}\ \emph {et~al.}(2019)\citenamefont
  {Hoferichter}, \citenamefont {Hoid},\ and\ \citenamefont
  {Kubis}}]{Hoferichter:2019gzf}%
  \BibitemOpen
  \bibfield  {author} {\bibinfo {author} {\bibfnamefont {M.}~\bibnamefont
  {Hoferichter}}, \bibinfo {author} {\bibfnamefont {B.-L.}\ \bibnamefont
  {Hoid}}, \ and\ \bibinfo {author} {\bibfnamefont {B.}~\bibnamefont {Kubis}},\
  }\href {\doibase 10.1007/JHEP08(2019)137} {\bibfield  {journal} {\bibinfo
  {journal} {JHEP}\ }\textbf {\bibinfo {volume} {08}},\ \bibinfo {pages} {137}
  (\bibinfo {year} {2019})},\ \Eprint {http://arxiv.org/abs/1907.01556}
  {arXiv:1907.01556 [hep-ph]} \BibitemShut {NoStop}%
\bibitem [{\citenamefont {Davier}\ \emph {et~al.}(2020)\citenamefont {Davier},
  \citenamefont {Hoecker}, \citenamefont {Malaescu},\ and\ \citenamefont
  {Zhang}}]{Davier:2019can}%
  \BibitemOpen
  \bibfield  {author} {\bibinfo {author} {\bibfnamefont {M.}~\bibnamefont
  {Davier}}, \bibinfo {author} {\bibfnamefont {A.}~\bibnamefont {Hoecker}},
  \bibinfo {author} {\bibfnamefont {B.}~\bibnamefont {Malaescu}}, \ and\
  \bibinfo {author} {\bibfnamefont {Z.}~\bibnamefont {Zhang}},\ }\href
  {\doibase 10.1140/epjc/s10052-020-7792-2} {\bibfield  {journal} {\bibinfo
  {journal} {Eur. Phys. J.}\ }\textbf {\bibinfo {volume} {C80}},\ \bibinfo
  {pages} {241} (\bibinfo {year} {2020})},\ \Eprint
  {http://arxiv.org/abs/1908.00921} {arXiv:1908.00921 [hep-ph]} \BibitemShut
  {NoStop}%
\bibitem [{\citenamefont {Keshavarzi}\ \emph {et~al.}(2020)\citenamefont
  {Keshavarzi}, \citenamefont {Nomura},\ and\ \citenamefont
  {Teubner}}]{Keshavarzi:2019abf}%
  \BibitemOpen
  \bibfield  {author} {\bibinfo {author} {\bibfnamefont {A.}~\bibnamefont
  {Keshavarzi}}, \bibinfo {author} {\bibfnamefont {D.}~\bibnamefont {Nomura}},
  \ and\ \bibinfo {author} {\bibfnamefont {T.}~\bibnamefont {Teubner}},\ }\href
  {\doibase 10.1103/PhysRevD.101.014029} {\bibfield  {journal} {\bibinfo
  {journal} {Phys. Rev.}\ }\textbf {\bibinfo {volume} {D101}},\ \bibinfo
  {pages} {014029} (\bibinfo {year} {2020})},\ \Eprint
  {http://arxiv.org/abs/1911.00367} {arXiv:1911.00367 [hep-ph]} \BibitemShut
  {NoStop}%
\bibitem [{\citenamefont {Kurz}\ \emph {et~al.}(2014)\citenamefont {Kurz},
  \citenamefont {Liu}, \citenamefont {Marquard},\ and\ \citenamefont
  {Steinhauser}}]{Kurz:2014wya}%
  \BibitemOpen
  \bibfield  {author} {\bibinfo {author} {\bibfnamefont {A.}~\bibnamefont
  {Kurz}}, \bibinfo {author} {\bibfnamefont {T.}~\bibnamefont {Liu}}, \bibinfo
  {author} {\bibfnamefont {P.}~\bibnamefont {Marquard}}, \ and\ \bibinfo
  {author} {\bibfnamefont {M.}~\bibnamefont {Steinhauser}},\ }\href {\doibase
  10.1016/j.physletb.2014.05.043} {\bibfield  {journal} {\bibinfo  {journal}
  {Phys. Lett.}\ }\textbf {\bibinfo {volume} {B734}},\ \bibinfo {pages} {144}
  (\bibinfo {year} {2014})},\ \Eprint {http://arxiv.org/abs/1403.6400}
  {arXiv:1403.6400 [hep-ph]} \BibitemShut {NoStop}%
\bibitem [{\citenamefont {Melnikov}\ and\ \citenamefont
  {Vainshtein}(2004)}]{Melnikov:2003xd}%
  \BibitemOpen
  \bibfield  {author} {\bibinfo {author} {\bibfnamefont {K.}~\bibnamefont
  {Melnikov}}\ and\ \bibinfo {author} {\bibfnamefont {A.}~\bibnamefont
  {Vainshtein}},\ }\href {\doibase 10.1103/PhysRevD.70.113006} {\bibfield
  {journal} {\bibinfo  {journal} {Phys. Rev.}\ }\textbf {\bibinfo {volume}
  {D70}},\ \bibinfo {pages} {113006} (\bibinfo {year} {2004})},\ \Eprint
  {http://arxiv.org/abs/hep-ph/0312226} {arXiv:hep-ph/0312226 [hep-ph]}
  \BibitemShut {NoStop}%
\bibitem [{\citenamefont {Masjuan}\ and\ \citenamefont
  {S{\'a}nchez-Puertas}(2017)}]{Masjuan:2017tvw}%
  \BibitemOpen
  \bibfield  {author} {\bibinfo {author} {\bibfnamefont {P.}~\bibnamefont
  {Masjuan}}\ and\ \bibinfo {author} {\bibfnamefont {P.}~\bibnamefont
  {S{\'a}nchez-Puertas}},\ }\href {\doibase 10.1103/PhysRevD.95.054026}
  {\bibfield  {journal} {\bibinfo  {journal} {Phys. Rev.}\ }\textbf {\bibinfo
  {volume} {D95}},\ \bibinfo {pages} {054026} (\bibinfo {year} {2017})},\
  \Eprint {http://arxiv.org/abs/1701.05829} {arXiv:1701.05829 [hep-ph]}
  \BibitemShut {NoStop}%
\bibitem [{\citenamefont {Colangelo}\ \emph {et~al.}(2017)\citenamefont
  {Colangelo}, \citenamefont {Hoferichter}, \citenamefont {Procura},\ and\
  \citenamefont {Stoffer}}]{Colangelo:2017fiz}%
  \BibitemOpen
  \bibfield  {author} {\bibinfo {author} {\bibfnamefont {G.}~\bibnamefont
  {Colangelo}}, \bibinfo {author} {\bibfnamefont {M.}~\bibnamefont
  {Hoferichter}}, \bibinfo {author} {\bibfnamefont {M.}~\bibnamefont
  {Procura}}, \ and\ \bibinfo {author} {\bibfnamefont {P.}~\bibnamefont
  {Stoffer}},\ }\href {\doibase 10.1007/JHEP04(2017)161} {\bibfield  {journal}
  {\bibinfo  {journal} {JHEP}\ }\textbf {\bibinfo {volume} {04}},\ \bibinfo
  {pages} {161} (\bibinfo {year} {2017})},\ \Eprint
  {http://arxiv.org/abs/1702.07347} {arXiv:1702.07347 [hep-ph]} \BibitemShut
  {NoStop}%
\bibitem [{\citenamefont {Hoferichter}\ \emph {et~al.}(2018)\citenamefont
  {Hoferichter}, \citenamefont {Hoid}, \citenamefont {Kubis}, \citenamefont
  {Leupold},\ and\ \citenamefont {Schneider}}]{Hoferichter:2018kwz}%
  \BibitemOpen
  \bibfield  {author} {\bibinfo {author} {\bibfnamefont {M.}~\bibnamefont
  {Hoferichter}}, \bibinfo {author} {\bibfnamefont {B.-L.}\ \bibnamefont
  {Hoid}}, \bibinfo {author} {\bibfnamefont {B.}~\bibnamefont {Kubis}},
  \bibinfo {author} {\bibfnamefont {S.}~\bibnamefont {Leupold}}, \ and\
  \bibinfo {author} {\bibfnamefont {S.~P.}\ \bibnamefont {Schneider}},\ }\href
  {\doibase 10.1007/JHEP10(2018)141} {\bibfield  {journal} {\bibinfo  {journal}
  {JHEP}\ }\textbf {\bibinfo {volume} {10}},\ \bibinfo {pages} {141} (\bibinfo
  {year} {2018})},\ \Eprint {http://arxiv.org/abs/1808.04823} {arXiv:1808.04823
  [hep-ph]} \BibitemShut {NoStop}%
\bibitem [{\citenamefont {G{\'e}rardin}\ \emph {et~al.}(2019)\citenamefont
  {G{\'e}rardin}, \citenamefont {Meyer},\ and\ \citenamefont
  {Nyffeler}}]{Gerardin:2019vio}%
  \BibitemOpen
  \bibfield  {author} {\bibinfo {author} {\bibfnamefont {A.}~\bibnamefont
  {G{\'e}rardin}}, \bibinfo {author} {\bibfnamefont {H.~B.}\ \bibnamefont
  {Meyer}}, \ and\ \bibinfo {author} {\bibfnamefont {A.}~\bibnamefont
  {Nyffeler}},\ }\href {\doibase 10.1103/PhysRevD.100.034520} {\bibfield
  {journal} {\bibinfo  {journal} {Phys. Rev.}\ }\textbf {\bibinfo {volume}
  {D100}},\ \bibinfo {pages} {034520} (\bibinfo {year} {2019})},\ \Eprint
  {http://arxiv.org/abs/1903.09471} {arXiv:1903.09471 [hep-lat]} \BibitemShut
  {NoStop}%
\bibitem [{\citenamefont {Bijnens}\ \emph {et~al.}(2019)\citenamefont
  {Bijnens}, \citenamefont {Hermansson-Truedsson},\ and\ \citenamefont
  {Rodr{\'i}guez-S{\'a}nchez}}]{Bijnens:2019ghy}%
  \BibitemOpen
  \bibfield  {author} {\bibinfo {author} {\bibfnamefont {J.}~\bibnamefont
  {Bijnens}}, \bibinfo {author} {\bibfnamefont {N.}~\bibnamefont
  {Hermansson-Truedsson}}, \ and\ \bibinfo {author} {\bibfnamefont
  {A.}~\bibnamefont {Rodr{\'i}guez-S{\'a}nchez}},\ }\href {\doibase
  10.1016/j.physletb.2019.134994} {\bibfield  {journal} {\bibinfo  {journal}
  {Phys. Lett.}\ }\textbf {\bibinfo {volume} {B798}},\ \bibinfo {pages}
  {134994} (\bibinfo {year} {2019})},\ \Eprint
  {http://arxiv.org/abs/1908.03331} {arXiv:1908.03331 [hep-ph]} \BibitemShut
  {NoStop}%
\bibitem [{\citenamefont {Colangelo}\ \emph {et~al.}(2020)\citenamefont
  {Colangelo}, \citenamefont {Hagelstein}, \citenamefont {Hoferichter},
  \citenamefont {Laub},\ and\ \citenamefont {Stoffer}}]{Colangelo:2019uex}%
  \BibitemOpen
  \bibfield  {author} {\bibinfo {author} {\bibfnamefont {G.}~\bibnamefont
  {Colangelo}}, \bibinfo {author} {\bibfnamefont {F.}~\bibnamefont
  {Hagelstein}}, \bibinfo {author} {\bibfnamefont {M.}~\bibnamefont
  {Hoferichter}}, \bibinfo {author} {\bibfnamefont {L.}~\bibnamefont {Laub}}, \
  and\ \bibinfo {author} {\bibfnamefont {P.}~\bibnamefont {Stoffer}},\ }\href
  {\doibase 10.1007/JHEP03(2020)101} {\bibfield  {journal} {\bibinfo  {journal}
  {JHEP}\ }\textbf {\bibinfo {volume} {03}},\ \bibinfo {pages} {101} (\bibinfo
  {year} {2020})},\ \Eprint {http://arxiv.org/abs/1910.13432} {arXiv:1910.13432
  [hep-ph]} \BibitemShut {NoStop}%
\bibitem [{\citenamefont {Blum}\ \emph {et~al.}(2020)\citenamefont {Blum},
  \citenamefont {Christ}, \citenamefont {Hayakawa}, \citenamefont {Izubuchi},
  \citenamefont {Jin}, \citenamefont {Jung},\ and\ \citenamefont
  {Lehner}}]{Blum:2019ugy}%
  \BibitemOpen
  \bibfield  {author} {\bibinfo {author} {\bibfnamefont {T.}~\bibnamefont
  {Blum}}, \bibinfo {author} {\bibfnamefont {N.}~\bibnamefont {Christ}},
  \bibinfo {author} {\bibfnamefont {M.}~\bibnamefont {Hayakawa}}, \bibinfo
  {author} {\bibfnamefont {T.}~\bibnamefont {Izubuchi}}, \bibinfo {author}
  {\bibfnamefont {L.}~\bibnamefont {Jin}}, \bibinfo {author} {\bibfnamefont
  {C.}~\bibnamefont {Jung}}, \ and\ \bibinfo {author} {\bibfnamefont
  {C.}~\bibnamefont {Lehner}},\ }\href {\doibase
  10.1103/PhysRevLett.124.132002} {\bibfield  {journal} {\bibinfo  {journal}
  {Phys. Rev. Lett.}\ }\textbf {\bibinfo {volume} {124}},\ \bibinfo {pages}
  {132002} (\bibinfo {year} {2020})},\ \Eprint
  {http://arxiv.org/abs/1911.08123} {arXiv:1911.08123 [hep-lat]} \BibitemShut
  {NoStop}%
\bibitem [{\citenamefont {Colangelo}\ \emph {et~al.}(2014)\citenamefont
  {Colangelo}, \citenamefont {Hoferichter}, \citenamefont {Nyffeler},
  \citenamefont {Passera},\ and\ \citenamefont {Stoffer}}]{Colangelo:2014qya}%
  \BibitemOpen
  \bibfield  {author} {\bibinfo {author} {\bibfnamefont {G.}~\bibnamefont
  {Colangelo}}, \bibinfo {author} {\bibfnamefont {M.}~\bibnamefont
  {Hoferichter}}, \bibinfo {author} {\bibfnamefont {A.}~\bibnamefont
  {Nyffeler}}, \bibinfo {author} {\bibfnamefont {M.}~\bibnamefont {Passera}}, \
  and\ \bibinfo {author} {\bibfnamefont {P.}~\bibnamefont {Stoffer}},\ }\href
  {\doibase 10.1016/j.physletb.2014.06.012} {\bibfield  {journal} {\bibinfo
  {journal} {Phys. Lett.}\ }\textbf {\bibinfo {volume} {B735}},\ \bibinfo
  {pages} {90} (\bibinfo {year} {2014})},\ \Eprint
  {http://arxiv.org/abs/1403.7512} {arXiv:1403.7512 [hep-ph]} \BibitemShut
  {NoStop}%
\bibitem [{\citenamefont {Fienberg}(2019)}]{fienberg2019status}%
  \BibitemOpen
  \bibfield  {author} {\bibinfo {author} {\bibfnamefont {A.~T.}\ \bibnamefont
  {Fienberg}},\ }\href@noop {} {\enquote {\bibinfo {title} {The status and
  prospects of the muon $g-2$ experiment at fermilab},}\ } (\bibinfo {year}
  {2019}),\ \Eprint {http://arxiv.org/abs/1905.05318} {arXiv:1905.05318
  [hep-ex]} \BibitemShut {NoStop}%
\bibitem [{\citenamefont {Sato}(2017)}]{Sato:2017sdn}%
  \BibitemOpen
  \bibfield  {author} {\bibinfo {author} {\bibfnamefont {Y.}~\bibnamefont
  {Sato}} (\bibinfo {collaboration} {E34}),\ }\href {\doibase
  10.22323/1.294.0006} {\bibfield  {journal} {\bibinfo  {journal} {PoS}\
  }\textbf {\bibinfo {volume} {KMI2017}},\ \bibinfo {pages} {006} (\bibinfo
  {year} {2017})}\BibitemShut {NoStop}%
\bibitem [{\citenamefont {Pospelov}(2009)}]{Pospelov_2009}%
  \BibitemOpen
  \bibfield  {author} {\bibinfo {author} {\bibfnamefont {M.}~\bibnamefont
  {Pospelov}},\ }\href {\doibase 10.1103/physrevd.80.095002} {\bibfield
  {journal} {\bibinfo  {journal} {Physical Review D}\ }\textbf {\bibinfo
  {volume} {80}} (\bibinfo {year} {2009}),\
  10.1103/physrevd.80.095002}\BibitemShut {NoStop}%
\bibitem [{\citenamefont {Freitas}\ \emph {et~al.}(2014)\citenamefont
  {Freitas}, \citenamefont {Lykken}, \citenamefont {Kell},\ and\ \citenamefont
  {Westhoff}}]{Freitas:2014pua}%
  \BibitemOpen
  \bibfield  {author} {\bibinfo {author} {\bibfnamefont {A.}~\bibnamefont
  {Freitas}}, \bibinfo {author} {\bibfnamefont {J.}~\bibnamefont {Lykken}},
  \bibinfo {author} {\bibfnamefont {S.}~\bibnamefont {Kell}}, \ and\ \bibinfo
  {author} {\bibfnamefont {S.}~\bibnamefont {Westhoff}},\ }\href {\doibase
  10.1007/JHEP09(2014)155} {\bibfield  {journal} {\bibinfo  {journal} {JHEP}\
  }\textbf {\bibinfo {volume} {05}},\ \bibinfo {pages} {145} (\bibinfo {year}
  {2014})},\ \bibinfo {note} {[Erratum: JHEP 09, 155 (2014)]},\ \Eprint
  {http://arxiv.org/abs/1402.7065} {arXiv:1402.7065 [hep-ph]} \BibitemShut
  {NoStop}%
\bibitem [{\citenamefont {Calibbi}\ \emph {et~al.}(2018)\citenamefont
  {Calibbi}, \citenamefont {Ziegler},\ and\ \citenamefont
  {Zupan}}]{Calibbi:2018rzv}%
  \BibitemOpen
  \bibfield  {author} {\bibinfo {author} {\bibfnamefont {L.}~\bibnamefont
  {Calibbi}}, \bibinfo {author} {\bibfnamefont {R.}~\bibnamefont {Ziegler}}, \
  and\ \bibinfo {author} {\bibfnamefont {J.}~\bibnamefont {Zupan}},\ }\href
  {\doibase 10.1007/JHEP07(2018)046} {\bibfield  {journal} {\bibinfo  {journal}
  {JHEP}\ }\textbf {\bibinfo {volume} {07}},\ \bibinfo {pages} {046} (\bibinfo
  {year} {2018})},\ \Eprint {http://arxiv.org/abs/1804.00009} {arXiv:1804.00009
  [hep-ph]} \BibitemShut {NoStop}%
\bibitem [{\citenamefont {Lindner}\ \emph {et~al.}(2018)\citenamefont
  {Lindner}, \citenamefont {Platscher},\ and\ \citenamefont
  {Queiroz}}]{Lindner:2016bgg}%
  \BibitemOpen
  \bibfield  {author} {\bibinfo {author} {\bibfnamefont {M.}~\bibnamefont
  {Lindner}}, \bibinfo {author} {\bibfnamefont {M.}~\bibnamefont {Platscher}},
  \ and\ \bibinfo {author} {\bibfnamefont {F.~S.}\ \bibnamefont {Queiroz}},\
  }\href {\doibase 10.1016/j.physrep.2017.12.001} {\bibfield  {journal}
  {\bibinfo  {journal} {Phys. Rept.}\ }\textbf {\bibinfo {volume} {731}},\
  \bibinfo {pages} {1} (\bibinfo {year} {2018})},\ \Eprint
  {http://arxiv.org/abs/1610.06587} {arXiv:1610.06587 [hep-ph]} \BibitemShut
  {NoStop}%
\bibitem [{\citenamefont {Kowalska}\ and\ \citenamefont
  {Sessolo}(2017)}]{Kowalska:2017iqv}%
  \BibitemOpen
  \bibfield  {author} {\bibinfo {author} {\bibfnamefont {K.}~\bibnamefont
  {Kowalska}}\ and\ \bibinfo {author} {\bibfnamefont {E.~M.}\ \bibnamefont
  {Sessolo}},\ }\href {\doibase 10.1007/JHEP09(2017)112} {\bibfield  {journal}
  {\bibinfo  {journal} {JHEP}\ }\textbf {\bibinfo {volume} {09}},\ \bibinfo
  {pages} {112} (\bibinfo {year} {2017})},\ \Eprint
  {http://arxiv.org/abs/1707.00753} {arXiv:1707.00753 [hep-ph]} \BibitemShut
  {NoStop}%
\bibitem [{\citenamefont {Barducci}\ \emph {et~al.}(2018)\citenamefont
  {Barducci}, \citenamefont {Deandrea}, \citenamefont {Moretti}, \citenamefont
  {Panizzi},\ and\ \citenamefont {Prager}}]{Barducci:2018esg}%
  \BibitemOpen
  \bibfield  {author} {\bibinfo {author} {\bibfnamefont {D.}~\bibnamefont
  {Barducci}}, \bibinfo {author} {\bibfnamefont {A.}~\bibnamefont {Deandrea}},
  \bibinfo {author} {\bibfnamefont {S.}~\bibnamefont {Moretti}}, \bibinfo
  {author} {\bibfnamefont {L.}~\bibnamefont {Panizzi}}, \ and\ \bibinfo
  {author} {\bibfnamefont {H.}~\bibnamefont {Prager}},\ }\href {\doibase
  10.1103/PhysRevD.97.075006} {\bibfield  {journal} {\bibinfo  {journal} {Phys.
  Rev. D}\ }\textbf {\bibinfo {volume} {97}},\ \bibinfo {pages} {075006}
  (\bibinfo {year} {2018})},\ \Eprint {http://arxiv.org/abs/1801.02707}
  {arXiv:1801.02707 [hep-ph]} \BibitemShut {NoStop}%
\bibitem [{\citenamefont {Kelso}\ \emph {et~al.}(2014)\citenamefont {Kelso},
  \citenamefont {Long}, \citenamefont {Martinez},\ and\ \citenamefont
  {Queiroz}}]{Kelso:2014qka}%
  \BibitemOpen
  \bibfield  {author} {\bibinfo {author} {\bibfnamefont {C.}~\bibnamefont
  {Kelso}}, \bibinfo {author} {\bibfnamefont {H.}~\bibnamefont {Long}},
  \bibinfo {author} {\bibfnamefont {R.}~\bibnamefont {Martinez}}, \ and\
  \bibinfo {author} {\bibfnamefont {F.~S.}\ \bibnamefont {Queiroz}},\ }\href
  {\doibase 10.1103/PhysRevD.90.113011} {\bibfield  {journal} {\bibinfo
  {journal} {Phys. Rev. D}\ }\textbf {\bibinfo {volume} {90}},\ \bibinfo
  {pages} {113011} (\bibinfo {year} {2014})},\ \Eprint
  {http://arxiv.org/abs/1408.6203} {arXiv:1408.6203 [hep-ph]} \BibitemShut
  {NoStop}%
\bibitem [{\citenamefont {Biggio}\ and\ \citenamefont
  {Bordone}(2015)}]{Biggio:2014ela}%
  \BibitemOpen
  \bibfield  {author} {\bibinfo {author} {\bibfnamefont {C.}~\bibnamefont
  {Biggio}}\ and\ \bibinfo {author} {\bibfnamefont {M.}~\bibnamefont
  {Bordone}},\ }\href {\doibase 10.1007/JHEP02(2015)099} {\bibfield  {journal}
  {\bibinfo  {journal} {JHEP}\ }\textbf {\bibinfo {volume} {02}},\ \bibinfo
  {pages} {099} (\bibinfo {year} {2015})},\ \Eprint
  {http://arxiv.org/abs/1411.6799} {arXiv:1411.6799 [hep-ph]} \BibitemShut
  {NoStop}%
\bibitem [{\citenamefont {Queiroz}\ and\ \citenamefont
  {Shepherd}(2014)}]{Queiroz:2014zfa}%
  \BibitemOpen
  \bibfield  {author} {\bibinfo {author} {\bibfnamefont {F.~S.}\ \bibnamefont
  {Queiroz}}\ and\ \bibinfo {author} {\bibfnamefont {W.}~\bibnamefont
  {Shepherd}},\ }\href {\doibase 10.1103/PhysRevD.89.095024} {\bibfield
  {journal} {\bibinfo  {journal} {Phys. Rev. D}\ }\textbf {\bibinfo {volume}
  {89}},\ \bibinfo {pages} {095024} (\bibinfo {year} {2014})},\ \Eprint
  {http://arxiv.org/abs/1403.2309} {arXiv:1403.2309 [hep-ph]} \BibitemShut
  {NoStop}%
\bibitem [{\citenamefont {Biggio}\ \emph {et~al.}(2016)\citenamefont {Biggio},
  \citenamefont {Bordone}, \citenamefont {Di~Luzio},\ and\ \citenamefont
  {Ridolfi}}]{Biggio:2016wyy}%
  \BibitemOpen
  \bibfield  {author} {\bibinfo {author} {\bibfnamefont {C.}~\bibnamefont
  {Biggio}}, \bibinfo {author} {\bibfnamefont {M.}~\bibnamefont {Bordone}},
  \bibinfo {author} {\bibfnamefont {L.}~\bibnamefont {Di~Luzio}}, \ and\
  \bibinfo {author} {\bibfnamefont {G.}~\bibnamefont {Ridolfi}},\ }\href
  {\doibase 10.1007/JHEP10(2016)002} {\bibfield  {journal} {\bibinfo  {journal}
  {JHEP}\ }\textbf {\bibinfo {volume} {10}},\ \bibinfo {pages} {002} (\bibinfo
  {year} {2016})},\ \Eprint {http://arxiv.org/abs/1607.07621} {arXiv:1607.07621
  [hep-ph]} \BibitemShut {NoStop}%
\bibitem [{\citenamefont {Agrawal}\ \emph {et~al.}(2014)\citenamefont
  {Agrawal}, \citenamefont {Chacko},\ and\ \citenamefont
  {Verhaaren}}]{Agrawal:2014ufa}%
  \BibitemOpen
  \bibfield  {author} {\bibinfo {author} {\bibfnamefont {P.}~\bibnamefont
  {Agrawal}}, \bibinfo {author} {\bibfnamefont {Z.}~\bibnamefont {Chacko}}, \
  and\ \bibinfo {author} {\bibfnamefont {C.~B.}\ \bibnamefont {Verhaaren}},\
  }\href {\doibase 10.1007/JHEP08(2014)147} {\bibfield  {journal} {\bibinfo
  {journal} {JHEP}\ }\textbf {\bibinfo {volume} {08}},\ \bibinfo {pages} {147}
  (\bibinfo {year} {2014})},\ \Eprint {http://arxiv.org/abs/1402.7369}
  {arXiv:1402.7369 [hep-ph]} \BibitemShut {NoStop}%
\bibitem [{\citenamefont {Gninenko}\ \emph {et~al.}(2015)\citenamefont
  {Gninenko}, \citenamefont {Krasnikov},\ and\ \citenamefont
  {Matveev}}]{Gninenko_2015}%
  \BibitemOpen
  \bibfield  {author} {\bibinfo {author} {\bibfnamefont {S.}~\bibnamefont
  {Gninenko}}, \bibinfo {author} {\bibfnamefont {N.}~\bibnamefont {Krasnikov}},
  \ and\ \bibinfo {author} {\bibfnamefont {V.}~\bibnamefont {Matveev}},\ }\href
  {\doibase 10.1103/physrevd.91.095015} {\bibfield  {journal} {\bibinfo
  {journal} {Physical Review D}\ }\textbf {\bibinfo {volume} {91}} (\bibinfo
  {year} {2015}),\ 10.1103/physrevd.91.095015}\BibitemShut {NoStop}%
\bibitem [{\citenamefont {Chen}\ \emph {et~al.}(2017)\citenamefont {Chen},
  \citenamefont {Pospelov},\ and\ \citenamefont {Zhong}}]{Chen_2017}%
  \BibitemOpen
  \bibfield  {author} {\bibinfo {author} {\bibfnamefont {C.-Y.}\ \bibnamefont
  {Chen}}, \bibinfo {author} {\bibfnamefont {M.}~\bibnamefont {Pospelov}}, \
  and\ \bibinfo {author} {\bibfnamefont {Y.-M.}\ \bibnamefont {Zhong}},\ }\href
  {\doibase 10.1103/physrevd.95.115005} {\bibfield  {journal} {\bibinfo
  {journal} {Physical Review D}\ }\textbf {\bibinfo {volume} {95}} (\bibinfo
  {year} {2017}),\ 10.1103/physrevd.95.115005}\BibitemShut {NoStop}%
\bibitem [{\citenamefont {Kahn}\ \emph {et~al.}(2018)\citenamefont {Kahn},
  \citenamefont {Krnjaic}, \citenamefont {Tran},\ and\ \citenamefont
  {Whitbeck}}]{Kahn_2018}%
  \BibitemOpen
  \bibfield  {author} {\bibinfo {author} {\bibfnamefont {Y.}~\bibnamefont
  {Kahn}}, \bibinfo {author} {\bibfnamefont {G.}~\bibnamefont {Krnjaic}},
  \bibinfo {author} {\bibfnamefont {N.}~\bibnamefont {Tran}}, \ and\ \bibinfo
  {author} {\bibfnamefont {A.}~\bibnamefont {Whitbeck}},\ }\href {\doibase
  10.1007/jhep09(2018)153} {\bibfield  {journal} {\bibinfo  {journal} {Journal
  of High Energy Physics}\ }\textbf {\bibinfo {volume} {2018}} (\bibinfo {year}
  {2018}),\ 10.1007/jhep09(2018)153}\BibitemShut {NoStop}%
\bibitem [{\citenamefont {Åkesson~et. al.}(2018)}]{kesson2018light}%
  \BibitemOpen
  \bibfield  {author} {\bibinfo {author} {\bibfnamefont {T.}~\bibnamefont
  {Åkesson~et. al.}},\ }\href@noop {} {\enquote {\bibinfo {title} {Light dark
  matter experiment (ldmx)},}\ } (\bibinfo {year} {2018}),\ \Eprint
  {http://arxiv.org/abs/1808.05219} {arXiv:1808.05219 [hep-ex]} \BibitemShut
  {NoStop}%
\bibitem [{\citenamefont {Berlin}\ \emph {et~al.}(2019)\citenamefont {Berlin},
  \citenamefont {Blinov}, \citenamefont {Krnjaic}, \citenamefont {Schuster},\
  and\ \citenamefont {Toro}}]{Berlin_2019}%
  \BibitemOpen
  \bibfield  {author} {\bibinfo {author} {\bibfnamefont {A.}~\bibnamefont
  {Berlin}}, \bibinfo {author} {\bibfnamefont {N.}~\bibnamefont {Blinov}},
  \bibinfo {author} {\bibfnamefont {G.}~\bibnamefont {Krnjaic}}, \bibinfo
  {author} {\bibfnamefont {P.}~\bibnamefont {Schuster}}, \ and\ \bibinfo
  {author} {\bibfnamefont {N.}~\bibnamefont {Toro}},\ }\href {\doibase
  10.1103/physrevd.99.075001} {\bibfield  {journal} {\bibinfo  {journal}
  {Physical Review D}\ }\textbf {\bibinfo {volume} {99}} (\bibinfo {year}
  {2019}),\ 10.1103/physrevd.99.075001}\BibitemShut {NoStop}%
\bibitem [{\citenamefont {Tsai}\ \emph {et~al.}(2019)\citenamefont {Tsai},
  \citenamefont {deNiverville},\ and\ \citenamefont {Liu}}]{Tsai:2019mtm}%
  \BibitemOpen
  \bibfield  {author} {\bibinfo {author} {\bibfnamefont {Y.-D.}\ \bibnamefont
  {Tsai}}, \bibinfo {author} {\bibfnamefont {P.}~\bibnamefont {deNiverville}},
  \ and\ \bibinfo {author} {\bibfnamefont {M.~X.}\ \bibnamefont {Liu}},\
  }\href@noop {} {\  (\bibinfo {year} {2019})},\ \Eprint
  {http://arxiv.org/abs/1908.07525} {arXiv:1908.07525 [hep-ph]} \BibitemShut
  {NoStop}%
\bibitem [{\citenamefont {Ballett}\ \emph {et~al.}(2019)\citenamefont
  {Ballett}, \citenamefont {Hostert}, \citenamefont {Pascoli}, \citenamefont
  {Perez-Gonzalez}, \citenamefont {Tabrizi},\ and\ \citenamefont
  {Funchal}}]{Ballett_2019}%
  \BibitemOpen
  \bibfield  {author} {\bibinfo {author} {\bibfnamefont {P.}~\bibnamefont
  {Ballett}}, \bibinfo {author} {\bibfnamefont {M.}~\bibnamefont {Hostert}},
  \bibinfo {author} {\bibfnamefont {S.}~\bibnamefont {Pascoli}}, \bibinfo
  {author} {\bibfnamefont {Y.~F.}\ \bibnamefont {Perez-Gonzalez}}, \bibinfo
  {author} {\bibfnamefont {Z.}~\bibnamefont {Tabrizi}}, \ and\ \bibinfo
  {author} {\bibfnamefont {R.~Z.}\ \bibnamefont {Funchal}},\ }\href {\doibase
  10.1007/jhep01(2019)119} {\bibfield  {journal} {\bibinfo  {journal} {Journal
  of High Energy Physics}\ }\textbf {\bibinfo {volume} {2019}} (\bibinfo {year}
  {2019}),\ 10.1007/jhep01(2019)119}\BibitemShut {NoStop}%
\bibitem [{\citenamefont {Mohlabeng}(2019)}]{Mohlabeng_2019}%
  \BibitemOpen
  \bibfield  {author} {\bibinfo {author} {\bibfnamefont {G.}~\bibnamefont
  {Mohlabeng}},\ }\href {\doibase 10.1103/physrevd.99.115001} {\bibfield
  {journal} {\bibinfo  {journal} {Physical Review D}\ }\textbf {\bibinfo
  {volume} {99}} (\bibinfo {year} {2019}),\
  10.1103/physrevd.99.115001}\BibitemShut {NoStop}%
\bibitem [{\citenamefont {Krnjaic}\ \emph
  {et~al.}(2020{\natexlab{a}})\citenamefont {Krnjaic}, \citenamefont
  {Marques-Tavares}, \citenamefont {Redigolo},\ and\ \citenamefont
  {Tobioka}}]{Krnjaic_2020}%
  \BibitemOpen
  \bibfield  {author} {\bibinfo {author} {\bibfnamefont {G.}~\bibnamefont
  {Krnjaic}}, \bibinfo {author} {\bibfnamefont {G.}~\bibnamefont
  {Marques-Tavares}}, \bibinfo {author} {\bibfnamefont {D.}~\bibnamefont
  {Redigolo}}, \ and\ \bibinfo {author} {\bibfnamefont {K.}~\bibnamefont
  {Tobioka}},\ }\href {\doibase 10.1103/physrevlett.124.041802} {\bibfield
  {journal} {\bibinfo  {journal} {Physical Review Letters}\ }\textbf {\bibinfo
  {volume} {124}} (\bibinfo {year} {2020}{\natexlab{a}}),\
  10.1103/physrevlett.124.041802}\BibitemShut {NoStop}%
\bibitem [{\citenamefont {Battaglieri}\ \emph {et~al.}(2017)\citenamefont
  {Battaglieri} \emph {et~al.}}]{Battaglieri:2017aum}%
  \BibitemOpen
  \bibfield  {author} {\bibinfo {author} {\bibfnamefont {M.}~\bibnamefont
  {Battaglieri}} \emph {et~al.},\ }in\ \href@noop {} {\emph {\bibinfo
  {booktitle} {{U.S. Cosmic Visions: New Ideas in Dark Matter}}}}\ (\bibinfo
  {year} {2017})\ \Eprint {http://arxiv.org/abs/1707.04591} {arXiv:1707.04591
  [hep-ph]} \BibitemShut {NoStop}%
\bibitem [{\citenamefont {Bauer}\ \emph {et~al.}(2018)\citenamefont {Bauer},
  \citenamefont {Foldenauer},\ and\ \citenamefont {Jaeckel}}]{Bauer_2018}%
  \BibitemOpen
  \bibfield  {author} {\bibinfo {author} {\bibfnamefont {M.}~\bibnamefont
  {Bauer}}, \bibinfo {author} {\bibfnamefont {P.}~\bibnamefont {Foldenauer}}, \
  and\ \bibinfo {author} {\bibfnamefont {J.}~\bibnamefont {Jaeckel}},\ }\href
  {\doibase 10.1007/jhep07(2018)094} {\bibfield  {journal} {\bibinfo  {journal}
  {Journal of High Energy Physics}\ }\textbf {\bibinfo {volume} {2018}}
  (\bibinfo {year} {2018}),\ 10.1007/jhep07(2018)094}\BibitemShut {NoStop}%
\bibitem [{\citenamefont {Lees}\ \emph {et~al.}(2016)\citenamefont {Lees} \emph
  {et~al.}}]{TheBABAR:2016rlg}%
  \BibitemOpen
  \bibfield  {author} {\bibinfo {author} {\bibfnamefont {J.}~\bibnamefont
  {Lees}} \emph {et~al.} (\bibinfo {collaboration} {BaBar}),\ }\href {\doibase
  10.1103/PhysRevD.94.011102} {\bibfield  {journal} {\bibinfo  {journal} {Phys.
  Rev. D}\ }\textbf {\bibinfo {volume} {94}},\ \bibinfo {pages} {011102}
  (\bibinfo {year} {2016})},\ \Eprint {http://arxiv.org/abs/1606.03501}
  {arXiv:1606.03501 [hep-ex]} \BibitemShut {NoStop}%
\bibitem [{\citenamefont {Escudero}\ \emph {et~al.}(2019)\citenamefont
  {Escudero}, \citenamefont {Hooper}, \citenamefont {Krnjaic},\ and\
  \citenamefont {Pierre}}]{Escudero:2019gzq}%
  \BibitemOpen
  \bibfield  {author} {\bibinfo {author} {\bibfnamefont {M.}~\bibnamefont
  {Escudero}}, \bibinfo {author} {\bibfnamefont {D.}~\bibnamefont {Hooper}},
  \bibinfo {author} {\bibfnamefont {G.}~\bibnamefont {Krnjaic}}, \ and\
  \bibinfo {author} {\bibfnamefont {M.}~\bibnamefont {Pierre}},\ }\href
  {\doibase 10.1007/JHEP03(2019)071} {\bibfield  {journal} {\bibinfo  {journal}
  {JHEP}\ }\textbf {\bibinfo {volume} {03}},\ \bibinfo {pages} {071} (\bibinfo
  {year} {2019})},\ \Eprint {http://arxiv.org/abs/1901.02010} {arXiv:1901.02010
  [hep-ph]} \BibitemShut {NoStop}%
\bibitem [{\citenamefont {Krnjaic}\ \emph
  {et~al.}(2020{\natexlab{b}})\citenamefont {Krnjaic}, \citenamefont
  {Marques-Tavares}, \citenamefont {Redigolo},\ and\ \citenamefont
  {Tobioka}}]{Krnjaic:2019rsv}%
  \BibitemOpen
  \bibfield  {author} {\bibinfo {author} {\bibfnamefont {G.}~\bibnamefont
  {Krnjaic}}, \bibinfo {author} {\bibfnamefont {G.}~\bibnamefont
  {Marques-Tavares}}, \bibinfo {author} {\bibfnamefont {D.}~\bibnamefont
  {Redigolo}}, \ and\ \bibinfo {author} {\bibfnamefont {K.}~\bibnamefont
  {Tobioka}},\ }\href {\doibase 10.1103/PhysRevLett.124.041802} {\bibfield
  {journal} {\bibinfo  {journal} {Phys. Rev. Lett.}\ }\textbf {\bibinfo
  {volume} {124}},\ \bibinfo {pages} {041802} (\bibinfo {year}
  {2020}{\natexlab{b}})},\ \Eprint {http://arxiv.org/abs/1902.07715}
  {arXiv:1902.07715 [hep-ph]} \BibitemShut {NoStop}%
\bibitem [{\citenamefont {Galon}\ \emph {et~al.}(2020)\citenamefont {Galon},
  \citenamefont {Kajamovitz}, \citenamefont {Shih}, \citenamefont {Soreq},\
  and\ \citenamefont {Tarem}}]{Galon:2019owl}%
  \BibitemOpen
  \bibfield  {author} {\bibinfo {author} {\bibfnamefont {I.}~\bibnamefont
  {Galon}}, \bibinfo {author} {\bibfnamefont {E.}~\bibnamefont {Kajamovitz}},
  \bibinfo {author} {\bibfnamefont {D.}~\bibnamefont {Shih}}, \bibinfo {author}
  {\bibfnamefont {Y.}~\bibnamefont {Soreq}}, \ and\ \bibinfo {author}
  {\bibfnamefont {S.}~\bibnamefont {Tarem}},\ }\href {\doibase
  10.1103/PhysRevD.101.011701} {\bibfield  {journal} {\bibinfo  {journal}
  {Phys. Rev. D}\ }\textbf {\bibinfo {volume} {101}},\ \bibinfo {pages}
  {011701} (\bibinfo {year} {2020})},\ \Eprint
  {http://arxiv.org/abs/1906.09272} {arXiv:1906.09272 [hep-ph]} \BibitemShut
  {NoStop}%
\bibitem [{\citenamefont {Janish}\ and\ \citenamefont
  {Ramani}(2020)}]{Janish:2020knz}%
  \BibitemOpen
  \bibfield  {author} {\bibinfo {author} {\bibfnamefont {R.}~\bibnamefont
  {Janish}}\ and\ \bibinfo {author} {\bibfnamefont {H.}~\bibnamefont
  {Ramani}},\ }\href@noop {} {\  (\bibinfo {year} {2020})},\ \Eprint
  {http://arxiv.org/abs/2006.10069} {arXiv:2006.10069 [hep-ph]} \BibitemShut
  {NoStop}%
\bibitem [{\citenamefont {Gninenko}\ \emph {et~al.}(2019)\citenamefont
  {Gninenko}, \citenamefont {Kirpichnikov}, \citenamefont {Kirsanov},\ and\
  \citenamefont {Krasnikov}}]{Gninenko:2019qiv}%
  \BibitemOpen
  \bibfield  {author} {\bibinfo {author} {\bibfnamefont {S.}~\bibnamefont
  {Gninenko}}, \bibinfo {author} {\bibfnamefont {D.}~\bibnamefont
  {Kirpichnikov}}, \bibinfo {author} {\bibfnamefont {M.}~\bibnamefont
  {Kirsanov}}, \ and\ \bibinfo {author} {\bibfnamefont {N.}~\bibnamefont
  {Krasnikov}},\ }\href {\doibase 10.1016/j.physletb.2019.07.015} {\bibfield
  {journal} {\bibinfo  {journal} {Phys. Lett. B}\ }\textbf {\bibinfo {volume}
  {796}},\ \bibinfo {pages} {117} (\bibinfo {year} {2019})},\ \Eprint
  {http://arxiv.org/abs/1903.07899} {arXiv:1903.07899 [hep-ph]} \BibitemShut
  {NoStop}%
\bibitem [{\citenamefont {Celentano}(2014)}]{Celentano:2014wya}%
  \BibitemOpen
  \bibfield  {author} {\bibinfo {author} {\bibfnamefont {A.}~\bibnamefont
  {Celentano}} (\bibinfo {collaboration} {HPS}),\ }\href {\doibase
  10.1088/1742-6596/556/1/012064} {\bibfield  {journal} {\bibinfo  {journal}
  {J. Phys. Conf. Ser.}\ }\textbf {\bibinfo {volume} {556}},\ \bibinfo {pages}
  {012064} (\bibinfo {year} {2014})},\ \Eprint
  {http://arxiv.org/abs/1505.02025} {arXiv:1505.02025 [physics.ins-det]}
  \BibitemShut {NoStop}%
\bibitem [{\citenamefont {Capdevilla}\ \emph {et~al.}()\citenamefont
  {Capdevilla}, \citenamefont {Curtin}, \citenamefont {Kahn},\ and\
  \citenamefont {Krnjaic}}]{muonbible}%
  \BibitemOpen
  \bibfield  {author} {\bibinfo {author} {\bibfnamefont {R.}~\bibnamefont
  {Capdevilla}}, \bibinfo {author} {\bibfnamefont {D.}~\bibnamefont {Curtin}},
  \bibinfo {author} {\bibfnamefont {Y.}~\bibnamefont {Kahn}}, \ and\ \bibinfo
  {author} {\bibfnamefont {G.}~\bibnamefont {Krnjaic}},\ }\href@noop {} {\
  }\Eprint {http://arxiv.org/abs/In preparation} {In preparation} \BibitemShut
  {NoStop}%
\bibitem [{\citenamefont {Calibbi}\ \emph {et~al.}(2019)\citenamefont
  {Calibbi}, \citenamefont {Li}, \citenamefont {Li},\ and\ \citenamefont
  {Zhu}}]{Calibbi:2019bay}%
  \BibitemOpen
  \bibfield  {author} {\bibinfo {author} {\bibfnamefont {L.}~\bibnamefont
  {Calibbi}}, \bibinfo {author} {\bibfnamefont {T.}~\bibnamefont {Li}},
  \bibinfo {author} {\bibfnamefont {Y.}~\bibnamefont {Li}}, \ and\ \bibinfo
  {author} {\bibfnamefont {B.}~\bibnamefont {Zhu}},\ }\href@noop {} {\
  (\bibinfo {year} {2019})},\ \Eprint {http://arxiv.org/abs/1912.02676}
  {arXiv:1912.02676 [hep-ph]} \BibitemShut {NoStop}%
\bibitem [{\citenamefont {Calibbi}\ \emph {et~al.}(2020)\citenamefont
  {Calibbi}, \citenamefont {L\'opez-Ib\'a\~nez}, \citenamefont {Melis},\ and\
  \citenamefont {Vives}}]{Calibbi:2020emz}%
  \BibitemOpen
  \bibfield  {author} {\bibinfo {author} {\bibfnamefont {L.}~\bibnamefont
  {Calibbi}}, \bibinfo {author} {\bibfnamefont {M.}~\bibnamefont
  {L\'opez-Ib\'a\~nez}}, \bibinfo {author} {\bibfnamefont {A.}~\bibnamefont
  {Melis}}, \ and\ \bibinfo {author} {\bibfnamefont {O.}~\bibnamefont
  {Vives}},\ }\href {\doibase 10.1007/JHEP06(2020)087} {\bibfield  {journal}
  {\bibinfo  {journal} {JHEP}\ }\textbf {\bibinfo {volume} {06}},\ \bibinfo
  {pages} {087} (\bibinfo {year} {2020})},\ \Eprint
  {http://arxiv.org/abs/2003.06633} {arXiv:2003.06633 [hep-ph]} \BibitemShut
  {NoStop}%
\bibitem [{\citenamefont {Crivellin}\ \emph {et~al.}(2018)\citenamefont
  {Crivellin}, \citenamefont {Hoferichter},\ and\ \citenamefont
  {Schmidt-Wellenburg}}]{Crivellin:2018qmi}%
  \BibitemOpen
  \bibfield  {author} {\bibinfo {author} {\bibfnamefont {A.}~\bibnamefont
  {Crivellin}}, \bibinfo {author} {\bibfnamefont {M.}~\bibnamefont
  {Hoferichter}}, \ and\ \bibinfo {author} {\bibfnamefont {P.}~\bibnamefont
  {Schmidt-Wellenburg}},\ }\href {\doibase 10.1103/PhysRevD.98.113002}
  {\bibfield  {journal} {\bibinfo  {journal} {Phys. Rev.}\ }\textbf {\bibinfo
  {volume} {D98}},\ \bibinfo {pages} {113002} (\bibinfo {year} {2018})},\
  \Eprint {http://arxiv.org/abs/1807.11484} {arXiv:1807.11484 [hep-ph]}
  \BibitemShut {NoStop}%
\bibitem [{\citenamefont {Lee}\ \emph {et~al.}(1977)\citenamefont {Lee},
  \citenamefont {Quigg},\ and\ \citenamefont {Thacker}}]{Lee:1977yc}%
  \BibitemOpen
  \bibfield  {author} {\bibinfo {author} {\bibfnamefont {B.~W.}\ \bibnamefont
  {Lee}}, \bibinfo {author} {\bibfnamefont {C.}~\bibnamefont {Quigg}}, \ and\
  \bibinfo {author} {\bibfnamefont {H.}~\bibnamefont {Thacker}},\ }\href
  {\doibase 10.1103/PhysRevLett.38.883} {\bibfield  {journal} {\bibinfo
  {journal} {Phys. Rev. Lett.}\ }\textbf {\bibinfo {volume} {38}},\ \bibinfo
  {pages} {883} (\bibinfo {year} {1977})}\BibitemShut {NoStop}%
\bibitem [{\citenamefont {Haber}(1994)}]{Haber:1994pe}%
  \BibitemOpen
  \bibfield  {author} {\bibinfo {author} {\bibfnamefont {H.~E.}\ \bibnamefont
  {Haber}},\ }in\ \href@noop {} {\emph {\bibinfo {booktitle} {{21st Annual SLAC
  Summer Institute on Particle Physics: Spin Structure in High-energy Processes
  (School: 26 Jul - 3 Aug, Topical Conference: 4-6 Aug) (SSI 93)}}}}\ (\bibinfo
  {year} {1994})\ pp.\ \bibinfo {pages} {231--272},\ \Eprint
  {http://arxiv.org/abs/hep-ph/9405376} {arXiv:hep-ph/9405376} \BibitemShut
  {NoStop}%
\bibitem [{\citenamefont {Castillo}\ \emph {et~al.}(2014)\citenamefont
  {Castillo}, \citenamefont {Diaz},\ and\ \citenamefont
  {Morales}}]{Castillo:2013uda}%
  \BibitemOpen
  \bibfield  {author} {\bibinfo {author} {\bibfnamefont {A.}~\bibnamefont
  {Castillo}}, \bibinfo {author} {\bibfnamefont {R.~A.}\ \bibnamefont {Diaz}},
  \ and\ \bibinfo {author} {\bibfnamefont {J.}~\bibnamefont {Morales}},\ }\href
  {\doibase 10.1142/S0217751X14500857} {\bibfield  {journal} {\bibinfo
  {journal} {Int. J. Mod. Phys. A}\ }\textbf {\bibinfo {volume} {29}},\
  \bibinfo {pages} {1450085} (\bibinfo {year} {2014})},\ \Eprint
  {http://arxiv.org/abs/1309.0831} {arXiv:1309.0831 [hep-ph]} \BibitemShut
  {NoStop}%
\bibitem [{\citenamefont {Goodsell}\ and\ \citenamefont
  {Staub}(2018)}]{Goodsell:2018tti}%
  \BibitemOpen
  \bibfield  {author} {\bibinfo {author} {\bibfnamefont {M.~D.}\ \bibnamefont
  {Goodsell}}\ and\ \bibinfo {author} {\bibfnamefont {F.}~\bibnamefont
  {Staub}},\ }\href {\doibase 10.1140/epjc/s10052-018-6127-z} {\bibfield
  {journal} {\bibinfo  {journal} {Eur. Phys. J. C}\ }\textbf {\bibinfo {volume}
  {78}},\ \bibinfo {pages} {649} (\bibinfo {year} {2018})},\ \Eprint
  {http://arxiv.org/abs/1805.07306} {arXiv:1805.07306 [hep-ph]} \BibitemShut
  {NoStop}%
\bibitem [{\citenamefont {Biondini}\ \emph {et~al.}(2019)\citenamefont
  {Biondini}, \citenamefont {Leonardi}, \citenamefont {Panella},\ and\
  \citenamefont {Presilla}}]{Biondini:2019tcc}%
  \BibitemOpen
  \bibfield  {author} {\bibinfo {author} {\bibfnamefont {S.}~\bibnamefont
  {Biondini}}, \bibinfo {author} {\bibfnamefont {R.}~\bibnamefont {Leonardi}},
  \bibinfo {author} {\bibfnamefont {O.}~\bibnamefont {Panella}}, \ and\
  \bibinfo {author} {\bibfnamefont {M.}~\bibnamefont {Presilla}},\ }\href
  {\doibase 10.1016/j.physletb.2019.06.042} {\bibfield  {journal} {\bibinfo
  {journal} {Phys. Lett. B}\ }\textbf {\bibinfo {volume} {795}},\ \bibinfo
  {pages} {644} (\bibinfo {year} {2019})},\ \bibinfo {note} {[Erratum:
  Phys.Lett.B 799, 134990 (2019)]},\ \Eprint {http://arxiv.org/abs/1903.12285}
  {arXiv:1903.12285 [hep-ph]} \BibitemShut {NoStop}%
\bibitem [{\citenamefont {Cesarotti}\ \emph {et~al.}(2019)\citenamefont
  {Cesarotti}, \citenamefont {Lu}, \citenamefont {Nakai}, \citenamefont
  {Parikh},\ and\ \citenamefont {Reece}}]{Cesarotti:2018huy}%
  \BibitemOpen
  \bibfield  {author} {\bibinfo {author} {\bibfnamefont {C.}~\bibnamefont
  {Cesarotti}}, \bibinfo {author} {\bibfnamefont {Q.}~\bibnamefont {Lu}},
  \bibinfo {author} {\bibfnamefont {Y.}~\bibnamefont {Nakai}}, \bibinfo
  {author} {\bibfnamefont {A.}~\bibnamefont {Parikh}}, \ and\ \bibinfo {author}
  {\bibfnamefont {M.}~\bibnamefont {Reece}},\ }\href {\doibase
  10.1007/JHEP05(2019)059} {\bibfield  {journal} {\bibinfo  {journal} {JHEP}\
  }\textbf {\bibinfo {volume} {05}},\ \bibinfo {pages} {059} (\bibinfo {year}
  {2019})},\ \Eprint {http://arxiv.org/abs/1810.07736} {arXiv:1810.07736
  [hep-ph]} \BibitemShut {NoStop}%
\bibitem [{\citenamefont {Giudice}\ and\ \citenamefont
  {Romanino}(2004)}]{Giudice:2004tc}%
  \BibitemOpen
  \bibfield  {author} {\bibinfo {author} {\bibfnamefont {G.}~\bibnamefont
  {Giudice}}\ and\ \bibinfo {author} {\bibfnamefont {A.}~\bibnamefont
  {Romanino}},\ }\href {\doibase 10.1016/j.nuclphysb.2004.08.001} {\bibfield
  {journal} {\bibinfo  {journal} {Nucl. Phys. B}\ }\textbf {\bibinfo {volume}
  {699}},\ \bibinfo {pages} {65} (\bibinfo {year} {2004})},\ \bibinfo {note}
  {[Erratum: Nucl.Phys.B 706, 487--487 (2005)]},\ \Eprint
  {http://arxiv.org/abs/hep-ph/0406088} {arXiv:hep-ph/0406088} \BibitemShut
  {NoStop}%
\bibitem [{\citenamefont {Arkani-Hamed}\ \emph {et~al.}(2005)\citenamefont
  {Arkani-Hamed}, \citenamefont {Dimopoulos}, \citenamefont {Giudice},\ and\
  \citenamefont {Romanino}}]{ArkaniHamed:2004yi}%
  \BibitemOpen
  \bibfield  {author} {\bibinfo {author} {\bibfnamefont {N.}~\bibnamefont
  {Arkani-Hamed}}, \bibinfo {author} {\bibfnamefont {S.}~\bibnamefont
  {Dimopoulos}}, \bibinfo {author} {\bibfnamefont {G.}~\bibnamefont {Giudice}},
  \ and\ \bibinfo {author} {\bibfnamefont {A.}~\bibnamefont {Romanino}},\
  }\href {\doibase 10.1016/j.nuclphysb.2004.12.026} {\bibfield  {journal}
  {\bibinfo  {journal} {Nucl. Phys. B}\ }\textbf {\bibinfo {volume} {709}},\
  \bibinfo {pages} {3} (\bibinfo {year} {2005})},\ \Eprint
  {http://arxiv.org/abs/hep-ph/0409232} {arXiv:hep-ph/0409232} \BibitemShut
  {NoStop}%
\bibitem [{\citenamefont {Delahaye}\ \emph {et~al.}(2019)\citenamefont
  {Delahaye}, \citenamefont {Diemoz}, \citenamefont {Long}, \citenamefont
  {Mansouli/'e}, \citenamefont {Pastrone}, \citenamefont {Rivkin},
  \citenamefont {Schulte}, \citenamefont {Skrinsky},\ and\ \citenamefont
  {Wulzer}}]{Delahaye:2019omf}%
  \BibitemOpen
  \bibfield  {author} {\bibinfo {author} {\bibfnamefont {J.~P.}\ \bibnamefont
  {Delahaye}}, \bibinfo {author} {\bibfnamefont {M.}~\bibnamefont {Diemoz}},
  \bibinfo {author} {\bibfnamefont {K.}~\bibnamefont {Long}}, \bibinfo {author}
  {\bibfnamefont {B.}~\bibnamefont {Mansouli/'e}}, \bibinfo {author}
  {\bibfnamefont {N.}~\bibnamefont {Pastrone}}, \bibinfo {author}
  {\bibfnamefont {L.}~\bibnamefont {Rivkin}}, \bibinfo {author} {\bibfnamefont
  {D.}~\bibnamefont {Schulte}}, \bibinfo {author} {\bibfnamefont
  {A.}~\bibnamefont {Skrinsky}}, \ and\ \bibinfo {author} {\bibfnamefont
  {A.}~\bibnamefont {Wulzer}},\ }\href@noop {} {\  (\bibinfo {year} {2019})},\
  \Eprint {http://arxiv.org/abs/1901.06150} {arXiv:1901.06150 [physics.acc-ph]}
  \BibitemShut {NoStop}%
\bibitem [{\citenamefont {Costantini}\ \emph {et~al.}(2020)\citenamefont
  {Costantini}, \citenamefont {De~Lillo}, \citenamefont {Maltoni},
  \citenamefont {Mantani}, \citenamefont {Mattelaer}, \citenamefont {Ruiz},\
  and\ \citenamefont {Zhao}}]{Costantini:2020stv}%
  \BibitemOpen
  \bibfield  {author} {\bibinfo {author} {\bibfnamefont {A.}~\bibnamefont
  {Costantini}}, \bibinfo {author} {\bibfnamefont {F.}~\bibnamefont
  {De~Lillo}}, \bibinfo {author} {\bibfnamefont {F.}~\bibnamefont {Maltoni}},
  \bibinfo {author} {\bibfnamefont {L.}~\bibnamefont {Mantani}}, \bibinfo
  {author} {\bibfnamefont {O.}~\bibnamefont {Mattelaer}}, \bibinfo {author}
  {\bibfnamefont {R.}~\bibnamefont {Ruiz}}, \ and\ \bibinfo {author}
  {\bibfnamefont {X.}~\bibnamefont {Zhao}}\ }(\bibinfo {year} {2020})\ \Eprint
  {http://arxiv.org/abs/2005.10289} {arXiv:2005.10289 [hep-ph]} \BibitemShut
  {NoStop}%
\bibitem [{\citenamefont {Alwall}\ \emph {et~al.}(2014)\citenamefont {Alwall},
  \citenamefont {Frederix}, \citenamefont {Frixione}, \citenamefont {Hirschi},
  \citenamefont {Maltoni}, \citenamefont {Mattelaer}, \citenamefont {Shao},
  \citenamefont {Stelzer}, \citenamefont {Torrielli},\ and\ \citenamefont
  {Zaro}}]{Alwall:2014hca}%
  \BibitemOpen
  \bibfield  {author} {\bibinfo {author} {\bibfnamefont {J.}~\bibnamefont
  {Alwall}}, \bibinfo {author} {\bibfnamefont {R.}~\bibnamefont {Frederix}},
  \bibinfo {author} {\bibfnamefont {S.}~\bibnamefont {Frixione}}, \bibinfo
  {author} {\bibfnamefont {V.}~\bibnamefont {Hirschi}}, \bibinfo {author}
  {\bibfnamefont {F.}~\bibnamefont {Maltoni}}, \bibinfo {author} {\bibfnamefont
  {O.}~\bibnamefont {Mattelaer}}, \bibinfo {author} {\bibfnamefont {H.~S.}\
  \bibnamefont {Shao}}, \bibinfo {author} {\bibfnamefont {T.}~\bibnamefont
  {Stelzer}}, \bibinfo {author} {\bibfnamefont {P.}~\bibnamefont {Torrielli}},
  \ and\ \bibinfo {author} {\bibfnamefont {M.}~\bibnamefont {Zaro}},\ }\href
  {\doibase 10.1007/JHEP07(2014)079} {\bibfield  {journal} {\bibinfo  {journal}
  {JHEP}\ }\textbf {\bibinfo {volume} {07}},\ \bibinfo {pages} {079} (\bibinfo
  {year} {2014})},\ \Eprint {http://arxiv.org/abs/1405.0301} {arXiv:1405.0301
  [hep-ph]} \BibitemShut {NoStop}%
\bibitem [{\citenamefont {Ellis}\ \emph {et~al.}(2019)\citenamefont {Ellis}
  \emph {et~al.}}]{Strategy:2019vxc}%
  \BibitemOpen
  \bibfield  {author} {\bibinfo {author} {\bibfnamefont {R.~K.}\ \bibnamefont
  {Ellis}} \emph {et~al.},\ }\href@noop {} {\  (\bibinfo {year} {2019})},\
  \Eprint {http://arxiv.org/abs/1910.11775} {arXiv:1910.11775 [hep-ex]}
  \BibitemShut {NoStop}%
\bibitem [{\citenamefont {Di~Luzio}\ \emph {et~al.}(2019)\citenamefont
  {Di~Luzio}, \citenamefont {Gröber},\ and\ \citenamefont
  {Panico}}]{DiLuzio:2018jwd}%
  \BibitemOpen
  \bibfield  {author} {\bibinfo {author} {\bibfnamefont {L.}~\bibnamefont
  {Di~Luzio}}, \bibinfo {author} {\bibfnamefont {R.}~\bibnamefont {Gröber}}, \
  and\ \bibinfo {author} {\bibfnamefont {G.}~\bibnamefont {Panico}},\ }\href
  {\doibase 10.1007/JHEP01(2019)011} {\bibfield  {journal} {\bibinfo  {journal}
  {JHEP}\ }\textbf {\bibinfo {volume} {01}},\ \bibinfo {pages} {011} (\bibinfo
  {year} {2019})},\ \Eprint {http://arxiv.org/abs/1810.10993} {arXiv:1810.10993
  [hep-ph]} \BibitemShut {NoStop}%
\bibitem [{\citenamefont {Chiesa}\ \emph {et~al.}(2020)\citenamefont {Chiesa},
  \citenamefont {Maltoni}, \citenamefont {Mantani}, \citenamefont {Mele},
  \citenamefont {Piccinini},\ and\ \citenamefont {Zhao}}]{chiesa2020measuring}%
  \BibitemOpen
  \bibfield  {author} {\bibinfo {author} {\bibfnamefont {M.}~\bibnamefont
  {Chiesa}}, \bibinfo {author} {\bibfnamefont {F.}~\bibnamefont {Maltoni}},
  \bibinfo {author} {\bibfnamefont {L.}~\bibnamefont {Mantani}}, \bibinfo
  {author} {\bibfnamefont {B.}~\bibnamefont {Mele}}, \bibinfo {author}
  {\bibfnamefont {F.}~\bibnamefont {Piccinini}}, \ and\ \bibinfo {author}
  {\bibfnamefont {X.}~\bibnamefont {Zhao}},\ }\href@noop {} {\enquote {\bibinfo
  {title} {Measuring the quartic higgs self-coupling at a multi-tev muon
  collider},}\ } (\bibinfo {year} {2020}),\ \Eprint
  {http://arxiv.org/abs/2003.13628} {arXiv:2003.13628 [hep-ph]} \BibitemShut
  {NoStop}%
\bibitem [{\citenamefont {Batell}\ \emph {et~al.}(2018)\citenamefont {Batell},
  \citenamefont {Freitas}, \citenamefont {Ismail},\ and\ \citenamefont
  {Mckeen}}]{Batell:2017kty}%
  \BibitemOpen
  \bibfield  {author} {\bibinfo {author} {\bibfnamefont {B.}~\bibnamefont
  {Batell}}, \bibinfo {author} {\bibfnamefont {A.}~\bibnamefont {Freitas}},
  \bibinfo {author} {\bibfnamefont {A.}~\bibnamefont {Ismail}}, \ and\ \bibinfo
  {author} {\bibfnamefont {D.}~\bibnamefont {Mckeen}},\ }\href {\doibase
  10.1103/PhysRevD.98.055026} {\bibfield  {journal} {\bibinfo  {journal} {Phys.
  Rev. D}\ }\textbf {\bibinfo {volume} {98}},\ \bibinfo {pages} {055026}
  (\bibinfo {year} {2018})},\ \Eprint {http://arxiv.org/abs/1712.10022}
  {arXiv:1712.10022 [hep-ph]} \BibitemShut {NoStop}%
\end{thebibliography}%

\end{document}